*Article*

# The P–T Probability Framework for Semantic Communication, Falsification, Confirmation, and Bayesian Reasoning

**Chenguang Lu**

survival99@gmail.com

**Graphical Abstract:**

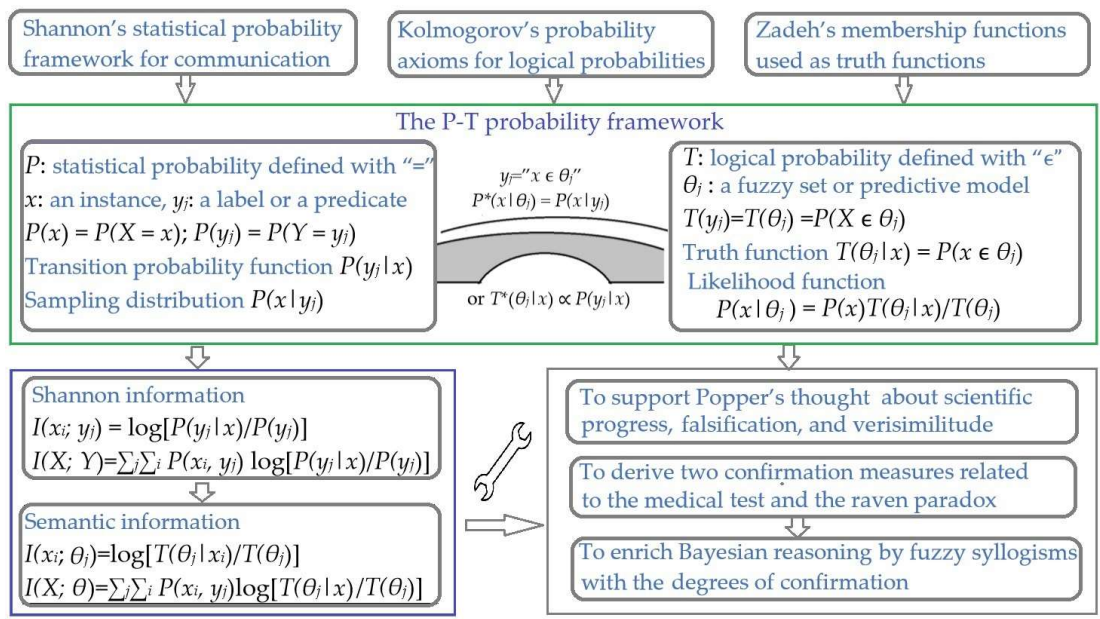

**Abstract:** Many researchers want to unify probability and logic by defining logical probability or probabilistic logic reasonably. This paper tries to unify statistics and logic so that we can use both statistical probability and logical probability at the same time. For this purpose, this paper proposes the P–T probability framework, which is assembled with Shannon's statistical probability framework for communication, Kolmogorov's probability axioms for logical probability, and Zadeh's membership functions used as truth functions. Two kinds of probabilities are connected by an extended Bayes' theorem, with which we can convert a likelihood function and a truth function from one to another. Hence, we can train truth functions (in logic) by sampling distributions (in statistics). This probability framework was developed in the author's long-term studies on semantic information, statistical learning, and color vision. This paper first proposes the P–T probability framework and explains different probabilities in it by its applications to semantic information theory. Then, this framework and the semantic information methods are applied to statistical learning, statistical mechanics, hypothesis evaluation (including falsification), confirmation, and Bayesian reasoning. Theoretical applications illustrate the reasonability and practicability of this framework. This framework is helpful for interpretable AI. To interpret neural networks, we need further study.

**1. Introduction**

*1.1. Are There Different Probabilities or One Probability with Different Interpretations?*

We can find several interpretations of probability: frequentist, logical, subjective, and propensity interpretations [1,2]. Almost every school believes that its interpretation is correct. However, we can see that different schools have their advantages. For example, Shannon's information theory [3] is a successful example of objectively frequentist probability. The maximum likelihood method [4] is an indispensable tool for statistical learning, which uses not only objective probabilities (sampling distributions) but also subjective probabilities (likelihood functions). Bayesians who developed Bayesian Inference [5] not only believe that probability is subjective, but also believe that predictive model $\theta$ is subjective and has a prior probability distribution. Bayesian Inference also has many successful applications. Using the propensity interpretation of probability [6], we can easily explain the objectivity of randomness in the physical world and the objectivity of the optimized likelihood function. As pointed out by Hájek [2], the above interpretations should be complementary; we need to "retain the distinct notions of physical logical/evidential, and subjective probability, with a rich tapestry of connections between them." The effort of this paper is in this direction.

The first question we confront in this direction is: Is there only one kind of probability that can be interpreted in different ways? Or, are there several kinds of different probabilities? In *The Logic of Scientific Discovery* ([7], pp. 252–258), Popper distinguishes the probability that a hypothesis is judged to be true and the probability that an event occurs. He thinks that the two are different. Even if we consider the probability of a hypothesis, we cannot only use "probability of statement", as Reichenbach does, to represent both its logical probability and its occurrence probability ([7], p. 252). Carnap [8] also distinguished probability 1 (logical probability) and probability 2 (statistical probability).

In reviewing the history of probability, Galavotti writes [1]:

> "Ever since its birth, probability has been characterized by a peculiar duality of meaning. As described by Hacking: probability is 'Janus faced. On the one side it is statistical, concerning itself with the stochastic laws of chance processes. On the other side it is epistemological, dedicated to assessing the reasonable degree of belief in propositions quite devoid of statistical background'".

Clearly, Popper and Carnap are not alone. I agree with Popper and Carnap that there exist two different probabilities: statistical probability and logical probability. I think that statistical probability may have frequentist, subjective, and propensity interpretations.

The statistical probability of a hypothesis or label is also the probability that a hypothesis or label is selected. We use an example to explain that a hypothesis has two different probabilities: selected probability and logical probability, at the same time.

**Example 1.** *There are five labels: $y_1$ = "child", $y_2$ = "youth", $y_3$ = "middle aged", $y_4$ = "elder", and $y_5$ = "adult". Notice that some youths and all middle-aged people and elders are also adults. Suppose that ten thousand people go through a door. For everyone denoted by $x$, entrance guards judge if $x$ is adult, or if "$x$ is adult" is true. If 7000 people are judged to be adults, then the logical probability of $y_5$ = "$x$ is adult" is 7000/10,000 = 0.7. If the* *task of entrance guards is to select one from the five labels for every person, there may be only 1000 people who are labeled "adult". The statistical probability of "adult" should be 1000/10,000 = 0.1.*

Why is the selected probability of $y_5$ is less than its logical probability? The reason is that other 6000 adults are labeled "youth", "middle aged", or "elder". In other words, the reason is that a person can only make one of five labels' selected probabilities increase 1/10,000. In contrast, a person can make two or more labels' logical probabilities increase 1/10,000. For example, a 20-year-old man can make both logical probabilities of "youth" and "adult" increase 1/10,000.

An extreme example is that the logical probability of a tautology, such as "*x* is adult or not adult", is 1. In contrast, its statistical probability is almost 0 in general, because a tautology is rarely selected.

From this example, we can find:

1. A hypothesis or label $y_j$ has two probabilities: a logical probability and a statistical or selected probability. If we use $P(y_j)$ to represent its statistical probability, we cannot use $P(y_j)$ for its logical probability.
2. Statistical probabilities are normalized (the sum is 1), whereas logical probabilities are not. The logical probability of a hypothesis is bigger than its statistical probability in general.
3. For given age *x*, such as *x* = 30, the sum of the truth values of five labels may be bigger than 1.
4. The logical probability of "adult" is related to the population age distribution $P(x)$, which is a statistical probability distribution. Clearly, the logical probabilities of "adult" obtained from the door of a school and the door of a hospital must be different.

The logical probability of $y_j$ calculated above accords with Reichenbach's frequentist definition of logical probability [9]. However, Reichenbach has not distinguished logical probability and statistical probability.

In the popular probability systems, such as the axiomatic system defined by Kolmogorov [10], all probabilities are represented by "*P*" so that we cannot use two kinds of different probabilities at the same time. Nor can we distinguish whether "*P*" represents a statistical probability or a logical probability, such as in popular confirmation measures [11].

Popper complained about Kolmogorov's probability system, saying: "he assumes that, in '*P*(*a*, *b*)' …a and b are sets; thereby excluding, among others, the logical interpretation according to which a and b are statements (or 'propositions', if you like)." ([7], pp. 330–331, where *P*(*a*, *b*) is a conditional probability and should be written as *P*(*a*|*b*) now.)

Popper hence proposed his own axiomatic system in *New Appendices of Conjectures and Refutations* ([7], pp. 329–355). In Popper's probability system, *a* and *b* in *P*(*a*|*b*) may be sets, predicates, propositions, and so on. However, Popper also only uses "*P*" to denote probability, and hence, his axiomatic system does not use statistical and logical probabilities at the same time. For this reason, Popper's probability system is not more practical than Kolmogorov's.

To distinguish statistical probability and logical probability and use both at the same time, I [12] propose the use of two different symbols "*P*" and "*T*" to identify statistical probabilities (including likelihood functions) and logical probabilities (including truth function). In my earlier papers [13–15], I used "*P*" for objective probabilities and "*Q*" for both subjective probability and logical probability. As subjective probability is also normalized and defined with "=" instead of "∈ (belong to)" (see Section 2.1), I still use "*P*" for subjective probability now.

*1.2. A Problem with the Extension of Logic: Can a Probability Function Be a Truth Function?*

Jaynes ([16], pp. 651–660) made a similar effort as Popper to improve Kolmogorov's probability system. He is a famous Bayesian, but he admits that we also need frequentist probability, such as the probability for sample distributions. He has proved that there is an equivalence between Boltzmann's entropy and Shannon's entropy, in which the probability is frequentist or objectively statistical. He summarizes that probability theory is the extension of logic. In his book [16], he uses many examples to show that much scientific reasoning is probabilistic. However, his probability system lacks truth functions that are multivalued or fuzzy, as reasoning tools.

A proposition has a truth value; all truth values of propositions with the same predicate and different subjects form a truth function. In the classical logic (binary logic), the characteristic function of a set is also the truth function of a hypothesis. Suppose that there is a predicate (or propositional function): $y_j(x)$ = "$x$ has property $a_j$", and set $A_j$ includes all $x$ that have property $a_j$. Then, the characteristic function of $A_j$ is the truth function of $y_j(x)$. For example, $x$ represents an age or a $x$-year-old person, and people with age ≥ 18 are defined as adults. Then, the set that includes all $x$ with property age ≥ 18 is $A_j$ = [18, ∞). Its characteristic function is also the truth function of predicate "$x$ is adult" (see Figure 1).

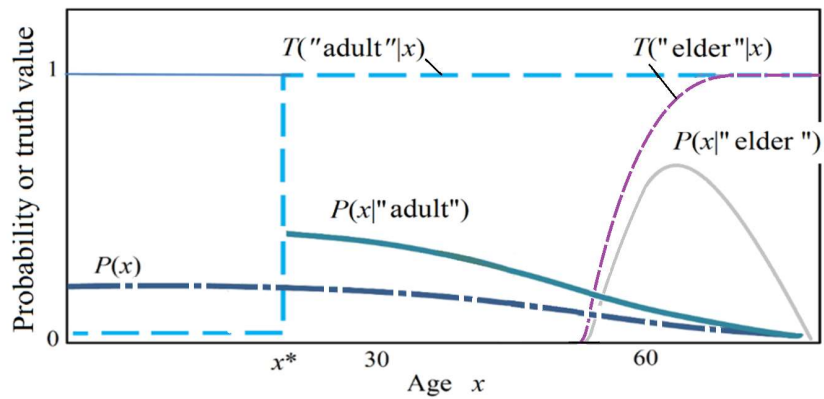

**Figure 1.** Solving the truth functions of "adult" and "elder" using prior and posterior probability distributions. The human brain can guess $T("adult"|x)$ (cyan dashed line) or the extension of "adult" from $P(x|"adult")$ (dark green line) and $P(x)$ (dark blue dashdotted line) and estimate $T("elder"|x)$ (purple thin dashed line) or the extension of "elder" from $P(x|"elder")$ (grey thin line) and $P(x)$. Equations (12) and (22) are new formulas for calculating $T("adult"|x)$ and $T("elder"|x)$.

If we extend binary logic to continuous value logic, the truth function should change between 0 and 1. Suppose that the number of a group of 60-year-old people is $N_{60}$, and $N_{60}{}^*$ people among the $N_{60}$ people are judged to be elderly. Then, the truth value of proposition "a 60-year-old person is elderly" is $N_{60}{}^*/N_{60}$. This truth value is about 0.8. If $x$ = 80, the truth value should be 1.

According to the logical interpretation of probability, a probability should be a continuous logical value, and a conditional probability function $P(y_j|x)$ with variable $x$ as the condition should be a truth function (e.g., fuzzy truth function). Shannon calls $P(y_j|x)$ the Transition Probability Function (TPF) ([3], p. 11). In the following, we use an example in natural language to explain why TPFs are not truth functions and how the human brain thinks using conceptual extensions (or denotations), which can be represented by truth functions.

**Example 2.** (refer to Figure 1). *Suppose we know the population age prior distribution $P(x)$ and the posterior distributions $P(x|"adult")$ and $P(x|"elder")$, where the extension of "adult" is crisp and that of "elder" is fuzzy. Please solve the truth functions or the extensions of labels "adult" and "elder".*

According to the existing probability theories, including Jaynes' probability theory, we cannot obtain the truth function from $P(x)$ and $P(x|y_j)$. We cannot get TPF $P(y_j|x)$ either. Because of $P(y_j|x)$ = $P(x|y_j)P(y_j)/P(x)$ according to Bayes' theorem, without $P(y_j)$, we cannot obtain $P(y_j|x)$. However, the human brain can estimate the extension of $y_j$ from $P(x)$ and $P(x|y_j)$ without $P(y_j)$ (see Figure 1). With the extension of a label, even if $P(x)$ is changed, the human brain can still predict the posterior distribution and classify people with different labels.

This example shows:

- The existing probability theories lack the methods used by the human brain (1) to find the extension of a label (an inductive method) and (2) to use the extension as the condition for reasoning or prediction (a deductive method).

- The TPF and the truth function are different, because a TPF is related to how many labels are used, whereas a truth function is not.

*1.3. Distinguishing (Fuzzy) Truth Values and Logical Probabilities—Using Zadeh's Fuzzy Set Theory*

Dubois and Prade [17] pointed out that there was a confusing tradition—researchers often confused the degree of belief (logical probability) in a hypothesis and the truth value of a proposition; we could only do logical operations with the truth values of propositions instead of logical probabilities. I agree with them. Section 7.2 will discuss why we cannot directly do logical operations with logical probabilities. Now, we further distinguish the logical probability and the (continuous or fuzzy) truth value.

The sentence "$x$ is elderly" (where $x$ is one of a group of people) is a propositional function; "this man is elderly" is a proposition. The former has a logical probability, whereas the latter has a truth value. As a propositional function includes a predicate "$x$ is elderly" and the universe $\mathcal{X}$, we also call a propositional function as a predicate. Therefore, we need to distinguish the logical probability of a predicate and the truth value of a proposition.

However, in the probability systems mentioned above, either we cannot distinguish the logical probability of a predicate and the truth value of a proposition (such as in Kolmogorov's probability system), or the truth value is binary (such as in the probability systems of Carnap [18] and Reichenbach [9]).

In Carnap's probability system [18], the logical probability of a predicate is irrelevant to the prior probability distribution $P(x)$. From Example 1, we can find that it is relevant.

Reichenbach [9] uses the statistical result of binary logic to define the logical probability of the logical expression of two proposition sequences (e.g., two predicates) without using continuous truth functions. Additionally, logical probability and statistical probability are not distinguished.

Therefore, the probability systems mentioned above are not satisfactory for the extension of logic. Fortunately, Zadeh's fuzzy set theory [19,20] provides more that we need than the above-mentioned probability systems. As the membership grade of an instance $x_i$ in fuzzy set $\theta_j$ is the truth value of proposition $y_j(x_i)$ = "$x_i$ is in $\theta_j$", the membership function of a fuzzy set is equivalent to the truth function of a predicate $y_j(x)$. The probability of a fuzzy event proposed by Zadeh [20] is just the logical probability of the predicate "$x$ is in $\theta_j$". Zadeh [21] thinks that the fuzzy set theory and the probability theory are complementary rather than competitive; fuzzy sets can enrich our concepts of probability.

However, it is still difficult for a machine to obtain truth functions or membership functions.

*1.4. Can We Use Sampling Distributions to Optimize Truth Functions or Membership Functions?*

Although the fuzzy set theory has made significant achievements in knowledge representation, fuzzy reasoning, and fuzzy control, it is not easy to use the fuzzy set theory for statistical learning, because membership functions are usually defined by experts rather than obtained from statistics.

A sample includes some examples. For instance, ($x$: age, $y$: label) is an example; many examples, such as (5, "child"), (20, "youth"), (70, "elder") …, form a sample. The posterior distribution $P(x|y_j)$ represents a sampling distribution. The core method of statistical learning is to optimize a likelihood function $P(x|\theta_j)$ ($\theta_j$ is a model or a set of parameters) with a sampling distribution $P(x|y_j)$. We also want to use sampling distributions to optimize truth functions or membership functions so that we can connect statistics and logic.

A significant advance in this direction is that Wang [22,23] developed the Random Set Falling Shadow Theory. According to this theory, a fuzzy set is produced by a random set; a set value taken by a random set is a range, such as ages 15–30 or 18–28 with label "youth". With many different ranges, we can calculate the proportion in which $x$ belongs to the random set. The limit of this proportion is the membership grade of $x$ in the corresponding fuzzy set.

However, it is still difficult to obtain membership functions or truth functions from the statistics of the random set. The cause is that there are a significant number of labeled examples with a single instance from real life, such as (70, "elder"), (60, "elder"), and (25, "youth"), but there are only very few examples with labeled sets or ranges, such as (15–30, "youth"). Therefore, the statistical method of random sets is not practical.

Due to the above reasons, fuzzy mathematics can perform well in expert systems, but not in statistical learning; many statisticians refuse to use fuzzy sets.

*1.5. Purpose, Methods, and Structure of This Paper*

Many researchers want to unify probability and logic. The natural idea is to define a logical probability system, as Keynes [24], Reichenbach [9], Popper [7], and Carnap [18] did. Some researchers use probabilities as logical variables to set up probability logic [25], such as two types of probability logic proposed by Reichenbach [9] and Adams [26]. Additionally, many researchers want to unify probabilities and fuzzy sets [27–29] following Zadeh. Many valuable efforts for the probabilification of the classical logic can be found in [25].

My effort is a little different. I also follow Zadeh to probabilify (e.g., fuzzify) the classical logic, but I want to use both statistical probabilities and logical probabilities at the same time and make them compatible, which means that the consequences of logical reasoning are equal to those of statistical reasoning when samples are enormous. In short, I want to unify statistics and logic, not only probability and logic.

A probability framework is a mathematical model that is constructed by the probabilities of some random variables taking values from some universes. For example, Shannon's communication model is a probability framework. The P–T probability framework includes statistical probabilities (represented by "P") and logical probability (represented by "T"). We may call Shannon's mathematical model for communication the P probability framework.

The purposes of this paper are to propose the P–T probability framework and test this framework through its uses in semantic communication, statistical learning, statistical mechanics, hypothesis evaluation (including falsification), confirmation, and Bayesian reasoning.

The main methods are

1. using the statistical probability framework, e.g., the P probability framework, adopted by Shannon for electrocommunication as the foundation, adding Kolmogorov's axioms (for logical probability) and Zadeh's membership functions (as truth functions) to the framework,
2. setting up the relationship between truth functions and likelihood functions by a new Bayes Theorem, called Bayes' Theorem III, so that we can optimize truth functions (in logic) with sampling distributions (in statistics), and
3. using the P–T probability framework and the semantic information Formulas (1) to express verisimilitude and testing severity and (2) to derive two practical confirmation measures and several new formulas to enrich Bayesian reasoning (including fuzzy syllogisms).

The P–T probability framework was gradually developed in my previous studies for the semantic information G theory (or say the G theory; "G" means the generalization of Shannon's information theory) [12,14,15] and statistical learning [12]. It is the first time in this paper that I name it the P–T probability framework and provide it with philosophical backgrounds and applications.

The rest of this paper is organized as follows. Section 2 defines the P–T probability framework. Section 3 explains this framework by its uses in semantic communication, statistical learning, and statistical mechanics. Section 4 shows how this framework and the semantic information measure support Popper's thought about scientific progress and hypothesis evaluation. Section 5 introduces two practical confirmation measures related to the medical test and the Raven Paradox, respectively. Section 6 summarizes various formulas for Bayesian reasoning and introduces a fuzzy logic that is compatible with Boolean Algebra. Section 7 discusses some questions, including how the theoretical applications exhibit the reasonability and practicality of the P–T probability framework. Section 8 ends with conclusions.

**2. The P–T Probability Framework**

*2.1. The Probability Framework Adopted by Shannon for Electrocommunication*

Shannon uses the frequentist probability system [30] for the P probability framework, which includes two random variables and two universes.

**Definition 1.** *(for the P probability framework adopted by Shannon):*

- $X$ is a discrete random variable taking a value $x \in \mathcal{X}$, where $\mathcal{X}$ is the universe $\{x_1, x_2, \ldots, x_m\}$; $P(x_i) = P(X = x_i)$ is the limit of the relative frequency of event $X = x_i$. In the following applications, $x$ represents an instance or a sample point.
- $Y$ is a discrete random variable taking a value $y \in \mathcal{Y} = \{y_1, y_2, \ldots, y_n\}$; $P(y_j) = P(Y = y_j)$. In the following applications, $y$ represents a label, hypothesis, or predicate.
- $P(y_j|x) = P(Y = y_j | X = x)$ is a Transition Probability Function (TPF) (named by Shannon [3]).

Shannon names $P(X)$ the source, $P(Y)$ the destination, and $P(Y|X)$ the channel (see Figure 2a). A Shannon channel is a transition probability matrix or a group of TPFs:

$$P(Y|X) \Leftrightarrow \begin{bmatrix} P(y_1|x_1) & P(y_1|x_2) & \ldots & P(y_1|x_m) \\ P(y_2|x_1) & P(y_2|x_2) & \ldots & P(y_2|x_m) \\ \ldots & \ldots & \ldots & \ldots \\ P(y_n|x_1) & P(y_n|x_2) & \ldots & P(y_n|x_m) \end{bmatrix} \Leftrightarrow \begin{bmatrix} P(y_1|x) \\ P(y_2|x) \\ \ldots \\ P(y_n|x) \end{bmatrix}, \quad (1)$$

where $\Leftrightarrow$ means equivalence. A Shannon channel or a TPF can be treated as a predictive model to produce the posterior distributions of $x$: $P(x|y_j)$, $j = 1, 2, \ldots, n$ (see Equation (6)).

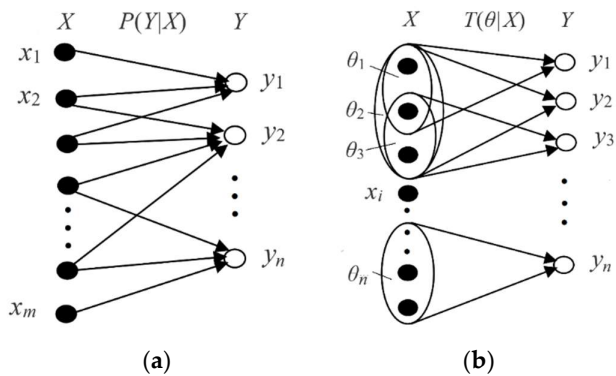

(a) (b)

**Figure 2.** The Shannon channel and the semantic channel. (**a**) The Shannon channel described by the P probability frame, (**b**) the semantic channel described by the P–T probability frame. A membership function ascertains the semantic meaning of $y_j$. A fuzzy set $\theta_j$ may be overlapped or included by another.

*2.2. The P–T Probability Framework for Semantic Communication*

**Definition 2.** *(for the P–T probability framework):*

- The $y_j$ is a label or a hypothesis, $y_j(x_i)$ is a proposition, and $y_j(x)$ is a propositional function. We also call $y_j$ or $y_j(x)$ a predicate. The $\theta_j$ is a fuzzy subset of universe $\mathcal{X}$, which is used to explain the semantic meaning of a propositional function $y_j(x) = $ "$x \in \theta_j$" = "$x$ belongs to $\theta_j$" = "$x$ is in $\theta_j$". The $\theta_j$ is also treated as a model or a set of model parameters.
- A probability that is defined with "=", such as $P(y_j) = P(Y = y_j)$, is a statistical probability. A probability that is defined with "$\in$", such as $P(X \in \theta_j)$, is a logical probability. To distinguish $P(Y = y_j)$ and $P(X \in \theta_j)$, we define $T(y_j) = T(\theta_j) = P(X \in \theta_j)$ as the logical probability of $y_j$.
- $T(y_j|x) = T(\theta_j|x) = P(x \in \theta_j) = P(X \in \theta_j | X = x)$ is the truth function of $y_j$ and the membership function of $\theta_j$. It changes between 0 and 1, and its maximum is 1.

According to Tarski's truth theory [31], $P(X \in \theta_j)$ is equivalent to $P(\text{"}X \in \theta\text{" is true}) = P(y_j \text{ is true})$. According to Davidson's truth condition semantics [32], the truth function of $y_j$ ascertains the (formally) semantic meaning of $y_j$. A group of truth functions, $T(\theta_j|x)$, $j$ = 1, 2, ..., $n$, forms a semantic channel:

$$T(\theta|X) \Leftrightarrow \begin{bmatrix} T(\theta_1|x_1) & T(\theta_1|x_2) & ... & T(\theta_1|x_m) \\ T(\theta_2|x_1) & T(\theta_2|x_2) & ... & T(\theta_2|x_m) \\ ... & ... & ... & ... \\ T(\theta_n|x_1) & T(\theta_n|x_2) & ... & T(\theta_n|x_m) \end{bmatrix} \Leftrightarrow \begin{bmatrix} T(\theta_1|x) \\ T(\theta_2|x) \\ ... \\ T(\theta_n|x) \end{bmatrix}. \qquad (2)$$

The Shannon channel and the semantic channel are illustrated in Figure 2.

Now, we explain the relationship between the above logical probability and Kolmogorov's probability. In Kolmogorov's probability system, a random variable takes a set as its value. Let $2^{\mathcal{X}}$ be the power set (Borel set), including all possible subsets of $\mathcal{X}$, and let $\Theta$ denote the random variable taking a value $\theta \in \{\theta_1, \theta_2, ..., \theta_{|2^{\mathcal{X}}|}\}$. Then, Kolmogorov's probability $P(\theta_j)$ is the probability of event $\Theta = \theta_j$. As $\Theta = \theta_j$ is equivalent to $X \in \theta_j$, we have $P(\Theta = \theta_j) = P(X \in \theta_j) = T(\theta_j)$. If all sets in $2^{\mathcal{X}}$ are crisp, then $T(\theta_j)$ becomes Kolmogorov's probability.

According to the above definition, we have the logical probability of $y_j$:

$$T(y_j) = T(\theta_j) = P(X \in \theta_j) = \sum_i P(x_i) T(\theta_j|x_i). \qquad (3)$$

The logical probability is the average truth value, and the truth value is the posterior logical probability. Notice that a proposition only has its truth value without its logical probability in general. For example, "every $x$ is white", where $x$ is a swan, has a logical probability, whereas "this swan is white" only has a truth value. Statistical probability $P(y_j)$ is equal to logical probability $T(\theta_j)$ for every $j$ only when the following two conditions are tenable:

- The universe of $\theta$ only contains some subsets of $2^{\mathcal{X}}$ that form a partition of $\mathcal{X}$, which means any two subsets in the universe are disjoint.
- The $y_j$ is always correctly selected.

For example, the universe of $Y$ is $\mathcal{Y}$ = {"Non-adult", "Adult"} or {"Child", "Youth", "Middle age", "Elder"}; $\mathcal{Y}$ ascertains a partition of $\mathcal{X}$. Further, if these labels are always correctly used, then $P(y_j) = T(\theta_j)$ for every $j$. Many researchers do not distinguish statistical probability and logical probability, because they suppose the above two conditions are always tenable. However, in real life, the two conditions are not tenable in general. When we use Kolmogorov's probability system for many applications without distinguishing statistical and logical probabilities, the two conditions are necessary. However, the latter condition is rarely mentioned.

The logical probability defined by Carnap and Bar-Hillel [33] is different from $T(\theta_j)$. Suppose there are three atomic propositions $a$, $b$, and $c$. The logical probability of a minimum term, such as $ab\bar{c}$ ($a$ and $b$ and not $c$), is 1/8; the logical probability of $b\bar{c} = ab\bar{c} \vee \bar{a}b\bar{c}$ is 1/4. This logical probability defined by Carnap and Bar-Hillel is irrelevant to the prior probability distribution $P(x)$. However, the logical probability $T(\theta_j)$ is related to $P(x)$. A less logical probability needs not only a smaller extension, but also rarer instances. For example, "$x$ is 20-year-old" has a less logical probability than "$x$ is young", but "$x$ is over 100 years old" with a larger extension has a still less logical probability than "$x$ is 20-year-old", because people over 100 are rare.

In mathematical logic, the true value of a universal predicate is equal to the logical multiplication of the truth values of all propositions with the predicate. Hence, Popper and Carnap conclude that the logical probability of a universal predicate is zero. This conclusion is bewildering. However, $T(\theta_j)$ is the logical probability of a predicate itself $y_j(x)$ itself without the quantifier (such as $\forall x$). It is the average truth value. Why do we define the logical probability in this way? One reason is that this mathematical definition conforms to the literal definition that logical probability is the probability in which a hypothesis is judged to be true. Another reason is that this definition is useful for extending Bayes' Theorem.

*2.3. Three Bayes' Theorems*

There are three Bayes' theorems, which are used by Bayes [34], Shannon [3], and me [12], respectively.

**Bayes' Theorem I.** *It was proposed by Bayes and can be expressed by Kolmogorov's probability. Suppose there are two sets A, B ∈ $2^x$; $A^c$ and $B^c$ are two complementary sets of A and B. Two symmetrical formulas express this theorem:*

$$T(B|A) = T(A|B)T(B)/T(A), \ T(A) = T(A|B)T(B) + T(A|B^c)T(B^c), \tag{4}$$

$$T(A|B) = T(B|A)T(A)/T(B), \ T(B) = T(B|A)T(A) + T(B|A^c)T(A^c). \tag{5}$$

Notice that there are one random variable, one universe, and two logical probabilities.

**Bayes' Theorem II.** *It is expressed by frequentist probabilities and used by Shannon. There are two symmetrical formulas:*

$$P(x|y_j) = P(y_j|x)P(x)/P(y_j), \ P(y_j) = \sum_i P(y_j|x_i)P(x_i), \tag{6}$$

$$P(y_j|x) = P(x|y_j)P(y_j)/P(x), \ P(x) = \sum_j P(x|y_j)P(y_j). \tag{7}$$

Notice there are two random variables, two universes, and two statistical probabilities.

**Bayes' Theorem III.** *It is used in the P–T probability framework. There are two asymmetrical formulas:*

$$P(x|\theta_j) = T(\theta_j|x)P(x)/T(\theta_j), \ T(\theta_j) = \sum_i P(x_i)T(\theta_j|x_i), \tag{8}$$

$$T(\theta_j|x) = T(\theta_j)P(x|\theta_j)/P(x), \ T(\theta_j) = 1/\max[P(x|\theta_j)/P(x)]. \tag{9}$$

The two formulas are asymmetrical, because there is a statistical probability and a logical probability on the left sides. Their proofs can be found in Appendix A.

$T(\theta_j)$ in Equation (8) is the horizontally normalizing constant (which makes the sum of $P(x|\theta_j)$ be 1), whereas $T(\theta_j)$ in Equation (9) is the longitudinally normalizing constant (which makes the maximum of $T(\theta_j|x)$ be 1).

We call $P(x|y_j)$ from Equation (6) the Bayes prediction, $P(x|\theta_j)$ from Equation (8) the semantic Bayes prediction, and Equation (8) the semantic Bayes formula. In [27], Dubois and Prade mentioned a formula, similar to Equation (8), proposed by S. F. Thomas and M. R. Civanlar earlier.

Equation (9) can be directly written as

$$T(\theta_j|x) = [P(x|\theta_j)/P(x)]/\max[P(x|\theta_j)/P(x)]. \tag{10}$$

*2.4. The Matching Relation between Statistical Probabilities and Logical Probabilities*

How can we understand a group of labels' extensions? We can learn from a sample.

Let **D** be a sample {(x(t), y(t))| t = 1 to N; x(t) ∈ 𝒳; y(t) ∈ 𝒴}, where (x(t), y(t)) is an example, and the use of each label is almost reasonable. All examples with label $y_j$ in **D** form a sub-sample denoted by **D**$_j$.

If **D**$_j$ is big enough, we can obtain smooth sampling distribution $P(x|y_j)$ from **D**$_j$. According to Fisher's maximum likelihood estimation, when $P(x|y_j) = P(x|\theta_j)$, we have the maximum likelihood between **D**$_j$ and $\theta_j$. Therefore, we set the matching relation between *P* and *T* by

$$P^*(x|\theta_j) = P(x|y_j), \ j = 1,2,\ldots,n, \tag{11}$$

where $P^*(x|\theta_j)$ is the optimized likelihood function. Then, we have the optimized truth functions:

$$T^*(\theta_j|x) = [P^*(x|\theta_j)/P(x)]/\max(P^*(x|\theta_j)/P(x)) = [P(x|y_j)/P(x)]/\max(P(x|y_j)/P(x)), \ j = 1, 2, \ldots, n. \tag{12}$$

Using Bayes' Theorem II, we further obtain

$$T^*(\theta_j|x) = P(y_j|x)/\max(P(y_j|x)), j = 1, 2, \ldots, n, \quad (13)$$

Which means that the semantic channel matches the Shannon channel. A TPF $P(y_j|x)$ indicates the using rule of a label $y_j$, and hence, the above Formulas (12) and (13) reflect Wittgenstein's thought: meaning lies in uses ([35], p. 80).

In my mind, "–" in "P–T" means Bayes' Theorem III and the above two formulas, which serves as a bridge between statistical probabilities and logical probabilities and hence as a bridge between statistics and logic.

Now, we can resolve the problem in Example 2 (see Figure 1). According to Equation (12), we have two optimized truth functions

$$T^*(\text{"adult"}|x) = [P(x|\text{"adult"})/P(x)]/\max(P(x|\text{"adult"})/P(x)),$$

$$T^*(\text{"elder"}|x) = [P(x|\text{"elder"})/P(x)]/\max(P(x|\text{"elder"})/P(x)),$$

no matter whether the extensions are fuzzy or not. After we obtain $T^*(\theta_j|x)$, we can make new probability predictions $P(x|\theta_j)$ using Bayes' Theorem III when $P(x)$ is changed.

In Dempster–Shafer theory, mass ≤ belief ≤ plausibility [36]. The above $P(y_j)$ is the mass, and $T(\theta_j)$ is like the belief. Suppose that $\mathcal{Y}$ has a subset $V_j = \{y_{j1}, y_{j2}, \ldots\}$, in which every label is not contradictory with $y_j$. We define $PL(V_j) = \sum_k P(y_{jk})$, which is like the plausibility. We also have $P(y_j) \le T(\theta_j) \le PL(V_j)$ and $P(y_j|x) \le T(\theta_j|x) \le PL(V_j|x)$.

Next, we prove that Equation (13) is compatible with the statistical method of random sets for membership functions proposed by Wang [23] (refer to Figure 3).

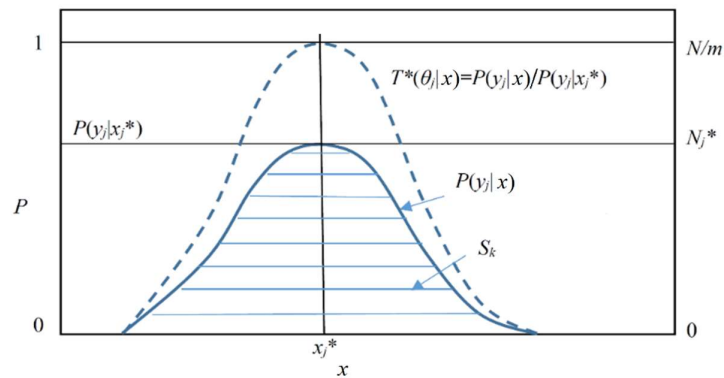

**Figure 3.** The optimized truth function (dashed line) is the same as the membership function obtained from the statistics of a random set [37]. $S_k$ is a set-value (a thin line); $N_j$ is the number of set-values.

Suppose that in sample **D**, $P(x)$ is flat, which means there are $N/m$ examples for every $x$. Then, the area under $P(y_j|x)$ is the number of examples in **D**$_j$. Suppose that example $(x_j^*, y_j)$ is most among examples $(x, y_j)$ with different $x$ in **D**$_j$, and the number of $(x_j^*, y_j)$ is $N_j^*$. Then, we divide all examples with $y_j$ into $N_j^*$ rows. Every row can be treated as a set $S_k$. It is easy to prove [37] that the truth function obtained from Equation (13) is the same as the membership function obtained from the statistics of a random set.

Therefore, the so obtained truth functions accord with Reichenbach's frequency interpretation of logical probability [9]. However, we should distinguish two kinds of probabilities that come from two different statistical methods. Popper was aware that the two kinds of probability were different, and there should be some relation between them ([7], pp. 252–258). Equation (13) should be what Popper wanted.

However, the above conclusions provided in Equations (12) and (13) need **D**$_j$ to be big enough. Otherwise, $P(y_j|x)$ is not smooth, and hence $P(x|\theta_j)$ is meaningless. In these cases, we need to use the maximum likelihood criterion or the maximum semantic information criterion to optimize truth functions (see Section 3.2).

*2.5. The Logical Probability and the Truth Function of a GPS Pointer or a Color Sense*

The semantic information is conveyed not only by natural languages, but also by clocks, scales, thermometers, GPS pointers, stock indexes, and so on, as pointed out by Floridi [38]. We can use a Gaussian truth function (a Gaussian function without normalizing coefficient)

$$T(\theta_j|x) = \exp[-|x - x_j|^2/(2\sigma^2)] \quad (14)$$

as the truth function of $y_j$ = "$x$ is about $x_j$," where $x_j$ is a reading, $x$ is the actual value, and $\sigma$ is the standard deviation. For a GPS device, $x_j$ is the pointed position (a vector) by $y_j$, $x$ is the actual position, and $\sigma$ is the Root Mean Square (RMS), which denotes the accuracy of a GPS device.

**Example 3.** *A GPS device is used in a train, and $P(x)$ is uniformly distributed on a line (see Figure 4). The GPS pointer has a deviation. Try to find the most possible position of the GPS device, according to $y_j$.*

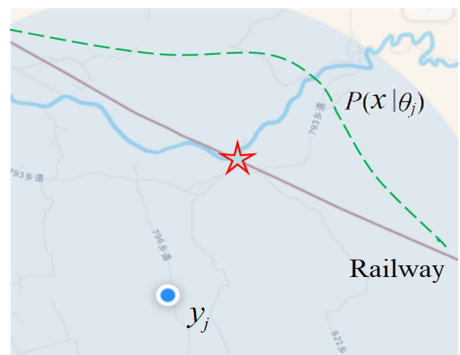

**Figure 4.** Illustrating a GPS device's positioning with a deviation. The round point is the pointed position, the star is the most possible position, and the dashed line is the predicted probability distribution. Source: author.

Using the statistical probability method, one might think that the GPS pointer and the circle for the RMS tell us a likelihood function $P(x|\theta_j)$ or a TPF $P(y_j|x)$. It is not a likelihood function, because the most possible position is not $x_j$ pointed by $y_j$. It is also not TPF $P(y_j|x)$, because we cannot know its maximum. It is reasonable to think that the GPS pointer provides a truth function.

Using the truth function, we can use Equation (8) to make a semantic Bayes prediction $P(x|\theta_j)$, according to which the position with the star is the most possible position. Most people can make the same prediction without using any mathematical formula. It seems that human brains automatically use a similar method: predicting according to the fuzzy extension of $y_j$ and the prior knowledge $P(x)$.

A color sense can also be treated as a reading of a device, which conveys semantic information [15]. In this case, the truth function of a color sense $y_j$ is also the similarity function or the confusion probability function between $x$ and $x_j$; logical probability $T(\theta_j)$ is the confusion probability of other colors that are confused with $x_j$ by our eyes.

We can also explain the logical probability of a hypothesis $y_j$ with the confusion probability. Suppose that there exists a Plato's idea $x_j$ for every fuzzy set $\theta_j$. Then, the membership function of $\theta_j$ or the truth function of $y_j$ is also the confusion probability function between $x$ and $x_j$ [14].

Like the semantic uncertainty of a GPS pointer, the semantic uncertainty of a microphysical quantity may also be expressed by a truth function instead of a probability function.

**3. The P–T Probability Framework for Semantic Communication, Statistical Learning, and Constraint Control**

*3.1. From Shannon's Information Measure to the Semantic Information Measure*

Shannon's mutual information is defined as:

$$I(X;Y)= \sum_j \sum_i P(x_i,y_j) \log \frac{P(x_i|y_j)}{P(x_i)}, \qquad (15)$$

where the base of the log is 2. If $Y = y_j$, $I(X; Y)$ becomes the Kullback–Leibler (KL) divergence:

$$I(X;y_j)= \sum_i P(x_i|y_j) \log \frac{P(x_i|y_j)}{P(x_i)}. \qquad (16)$$

If $X = x_i$, $I(X; y_j)$ becomes

$$I(x_i;y_j)= \log \frac{P(x_i|y_j)}{P(x_i)} = \log \frac{P(y_j|x_i)}{P(y_j)}. \qquad (17)$$

Using likelihood function $P(x|\theta_j)$ to replace posterior distribution $P(x|y_j)$, we have (the amount of) semantic information conveyed by $y_j$ about $x_i$:

$$I(x_i;\theta_j) = \log \frac{P(x_i|\theta_j)}{P(x_i)} = \log \frac{T(\theta_j|x_i)}{T(\theta_j)}. \qquad (18)$$

The above formula makes use of Bayes' Theorem III. Its philosophical meaning will be explained in Section 4. If the truth value of proposition $y_j(x_i)$ is always 1, then the above formula becomes Carnap and Bar-Hillel's semantic information formula [33].

The above semantic information formula can be used to measure not only the semantic information of natural languages, but also the semantic information of a GPS pointer, a thermometer, or a color sense. In the latter cases, the truth function also means the similarity function or the confusion probability function.

To average $I(x_i; \theta_j)$ for different $x_i$, we have average semantic information

$$I(X;\theta_j) = \sum_i P(x_i|y_j) \log \frac{P(x_i|\theta_j)}{P(x_i)} = \sum_i P(x_i|y_j) \log \frac{T(\theta_j|x_i)}{T(\theta_j)}, \qquad (19)$$

where $P(x_i|y_j)$ ($i = 1,2, …$) is the sampling distribution. This formula can be used to optimize truth functions.

To average $I(X; \theta_j)$ for different $y_j$, we have semantic mutual information:

$$I(X;\Theta) = \sum_j P(y_j) \sum_i P(x_i|y_j) \log \frac{P(x_i|\theta_j)}{P(x_i)} = \sum_i \sum_j P(x_i) P(y_j|x_i) \log \frac{T(\theta_j|x_i)}{T(\theta_j)}. \qquad (20)$$

Both $I(X; \theta_j)$ and $I(X; \Theta)$ can be used as the criterion of classifications. We also call $I(X; \theta_j)$ generalized Kullback–Leibler (KL) information and $I(X; \Theta)$ generalized mutual information.

If the truth function is a Gaussian truth function, $I(X; \Theta)$ is equal to the generalized entropy minus the mean relative squared error:

$$I(X;\Theta) = -\sum_j P(y_j) \log T(\theta_j) - \sum_i \sum_j P(x_i,y_j)(x_i - y_j)^2 / (2\sigma^2). \qquad (21)$$

Therefore, the maximum semantic mutual information criterion is similar to the Regularized Least Square (RLS) criterion that is getting popular in machine learning.

Shannon proposed the fidelity evaluation function for data compression in his famous paper [3]. A specific fidelity evaluation function is the rate-distortion function $R(D)$ [39], where $D$ is the upper limit of the average of distortion $d(x_i, y_j)$ between $x_i$ and $y_j$, and $R$ is the minimum mutual information for given $D$. $R(D)$ will be further introduced in relation to random events' control in Section 3.3.

Replacing $d(x_i, y_j)$ with $I(x_i; \theta_j)$, I developed another fidelity evaluation function $R(G)$ [12,15], where $G$ is the lower limit of semantic mutual information $I(X; \Theta)$, and $R(G)$ is the minimum Shannon mutual information for given $G$. $G/R(G)$ indicates the communication efficiency, whose upper limit is 1 when $P(x|\theta_j) = P(x|y_j)$ for all $j$. $R(G)$ is called the rate-verisimilitude function (this verisimilitude will be further discussed in Section 4.3). The rate-verisimilitude function is useful for data compression according to visual discrimination [15] and the convergence proofs of mixture models and maximum mutual information classifications [12].

### 3.2. Optimizing Truth Functions and Classifications for Natural Language

An essential task of statistical learning is to optimize predictive models (such as likelihood functions and logistic functions) and classifications. As by using a truth function $T(\theta_j|x)$ and the prior probability distribution $P(x)$ we can produce a likelihood function; a truth function can also be treated as a predictive model. Additionally, a truth function as a predictive model has the advantage that it still works when $P(x)$ is changed.

Now, we consider Example 2 again. When sample $\mathbf{D}_j$ is not big enough, and hence $P(x|y_j)$ is unsmooth, we cannot use Equations (12) or (13) to obtain a smooth truth function. In this case, we can use the generalized KL formula to get an optimized continuous truth function. For example, the optimized truth function is

$$T^*(\theta_{\text{elder}} | x) = \arg\max_{\theta_{\text{elder}}} \sum_i P(x_i | \text{"elder"}) \log \frac{T(\theta_{\text{elder}} | x_i)}{T(\theta_{\text{elder}})}. \tag{22}$$

where "arg max $\sum$" means to maximize the sum by selecting parameters $\theta_{\text{elder}}$ of the truth function. In this case, we can assume that $T(\theta_{\text{elder}})$ is a logistic function: $T(\theta_{\text{elder}}) = 1/[1 + \exp(-u(x - v))]$, where $u$ and $v$ are two parameters to be optimized. If we know $P(y_j|x)$ without knowing $P(x)$, we may assume that $P(x)$ is constant to obtain sampling distribution $P(x|y_j)$ and logical probability $T(\theta_{\text{elder}})$ [12].

It is called Logical Bayesian Inference [12] to (1) obtain the optimized truth function $T^*(\theta_j|x)$ from sampling distributions $P(x|y_j)$ and $P(x)$ and (2) make the semantic Bayes prediction $P(x|\theta_j)$ using $T^*(\theta_j|x)$ and $P(x)$. Logical Bayesian Inference is different from Bayesian Inference [5]. The former uses prior $P(x)$, whereas the latter uses prior $P(\theta)$.

In statistical learning, it is called label learning to obtain smooth truth functions or membership functions with parameters. From the philosophical perspective, it is the induction of labels' extensions.

In popular statistical learning methods, the label learning of two complementary labels, such as "elder" and "non-elder", is easy, because we can use a pair of logistic functions as two TPFs $P(\theta_1|x)$ and $P(\theta_0|x)$ or two truth functions $T(\theta_1|x)$ and $T(\theta_0|x)$ with parameters. However, multi-label learning is difficult [40], because it is impossible to design $n$ TPFs with parameters. Nevertheless, using the P–T probability framework, multi-label learning is also easy, because every label's learning is independent [12].

With optimized truth functions, we can classify different instances into different classes using a classifier $y_j = f(x)$. For instance, we can classify people with different ages into classes with labels "child", "youth", "adult", "middle aged", and "elder". Using the maximum semantic information criterion, the classifier is

$$y_j^* = f(x) = \arg\max_{y_j} \log I(x; \theta_j) = \arg\max_{y_j} \log[T(\theta_j | x) / T(\theta_j)]. \tag{23}$$

This classifier changes with the population age distribution $P(x)$. With the prolongation of the human life span, $v$ in the truth function $T^*(\theta_{\text{elder}}|x)$ and the division point of "elder" will automatically increase [12].

The maximum semantic information criterion is compatible with the maximum likelihood criterion and is different from the maximum correctness criterion. It encourages us to reduce the underreports of small probability events. For example, if we classify people over 60 into the elder class according to the maximum correctness criterion, then we may classify people over 58 into the elder class according to the maximum semantic information criterion, because the elder people are less numerous than non-elder or the middle-aged people. To predict earthquakes, if one uses the maximum correctness criterion, he may always predict "no earthquake will happen next week". Such predictions are meaningless. If one uses the maximum semantic information criterion, he will predict "the earthquake will be possible next week" in some cases, even if he will make more mistakes.

The P–T probability framework and the G theory can also be used for improving maximum mutual information classifications and mixture models [12].

### 3.3. Truth Functions Used as Distribution Constraint Functions for Random Events' Control

A truth function is also a membership function and can be treated as the constraint function of the probability distribution of a random event or the density distribution of random particles. We simply call this function the Distribution Constraint Function (DCF).

The KL divergence and the Shannon mutual information can also be used to measure the control amount of controlling a random event. Let $X$ be a random event, and $P(x) = P(X = x)$ be the prior probability distribution. $X$ may be the energy of a gas molecule, the size of a machined part, the age of one of the people in a country, and so on. $P(x)$ may also be a density function.

If $P(x|y_j)$ is the posterior distribution after a control action $y_j$, then the KL divergence $I(X; y_j)$ is the control amount (in bits), which reflects the complexity of the control. If the ideal posterior distribution is $P(x|\theta_j)$, then the effective control amount is

$$I_c(X;\theta_j) = \sum_i P(x_i|\theta_j) \log \frac{P(x_i|y_j)}{P(x_i)}. \tag{24}$$

Notice that $P(x_i|\theta_j)$ is on the left of "log" instead of the right. For generalized KL information $I(X; \theta_j)$, when prediction $P(x|\theta_j)$ approaches fact $P(x|y_j)$, $I(X; \theta_j)$ approaches its maximum. In contrast, for effective control amount $I_c(X; \theta_j)$, as fact $P(x|y_j)$ approaches ideality $P(x|\theta_j)$, $I_c(X; \theta_j)$ approaches its maximum. For an inadequate $P(x|y_j)$, $I_c(X; \theta_j)$ may be negative. $P(x|y_j)$ may also have parameters.

A truth function $T(\theta_j|x)$ used as a DCF means that there should be

$$1 - P(x|y_j) \le 1 - P(x|\theta_j) = 1 - P(x)T(\theta_j|x)/T(\theta_j), \text{ for } T(\theta_j|x) < 1. \tag{25}$$

If $\theta_j$ is a crisp set, this condition means that $x$ cannot be outside of $\theta_j$. If $\theta_j$ is fuzzy, it means that $x$ outside of $\theta_j$ should be limited. There are many distributions $P(x|y_j)$ that meet the above condition, but only one needs the minimum KL information $I(X; y_j)$. For example, assuming that $x_j$ makes $T(\theta_j|x) = 1$, if $P(x|y_j) = 1$ for $x = x_j$ and $P(x|y_j) = 0$ for $x \ne x_j$, then $P(x|y_j)$ meets the above condition. However, this $P(x|y_j)$ needs information $I(X; y_j)$ that is not minimum.

I studied the rate-tolerance function $R(\Theta)$ [41] that is the extension of the complexity-distortion function $R(C)$ [42]. Unlike the rate-distortion function, the constraint condition of the complexity-distortion function is that every distortion $d(x_i, y_j) = (x_i - y_j)^2$ is less than a given value $C$, which means the constraint sets possess the same magnitude. Unlike the constraint condition of $R(C)$, the constraint condition of $R(\Theta)$ is that the constraint sets are fuzzy and possess different magnitudes. I have concluded [14,15]:

- For given DCFs $T(\theta_j|x)$ ($j = 1, 2, …, n$) and $P(x)$, when $P(x|y_j) = P(x|\theta_j) = P(x)T(\theta_j|x)/T(\theta_j)$, the KL divergence $I(X; y_j)$ and Shannon's mutual information $I(X; Y)$ reach their minima; the effective control amount $I_c(X; y_j)$ reaches its maximum. If every set $\theta_j$ is crisp, $I(X; y_j) = -\log T(\theta_j)$ and $I(X; Y) = -\sum_j P(y_j)\log T(\theta_j)$.
- A rate-distortion function $R(D)$ is equivalent to a rate-tolerance function $R(\Theta)$, and a semantic mutual information formula can express it with truth functions or DCFs (see Appendix B for details). However, an $R(\Theta)$ function may not be equivalent to an $R(D)$ function, and hence, $R(D)$ is a special case of $R(\Theta)$.

Let $d_{ij} = d(x_i, y_j)$ be the distortion or the loss when we use $y_j$ to represent $x_i$. $D$ is the upper limit of the average distortion. For given $P(x)$, we can obtain the minimum Shannon mutual information $I(X; Y)$, e.g., $R(D)$. The parameterization of $R(D)$ ([43], P. 32) includes two formulas:

$$D(s) = \sum_i \sum_j d_{ij} P(x_i) P(y_j) \exp(sd_{ij}) / \lambda_i,$$
$$R(s) = sD(s) - \sum_i P(x_i) \ln \lambda_i, \quad \lambda_i = \sum_j P(y_j) \exp(sd_{ij}), \quad (26)$$

where parameter $s = dR/dD \leq 0$ reflects the slope of $R(D)$. The posterior distribution $P(y|x_i)$ is

$$P(y|x_i) = P(y) \exp(sd_{ij}) / \lambda_i, \quad (27)$$

which makes $I(X; Y)$ reach its minimum.

As $s \leq 0$, the maximum of $\exp(sd_{ij})$ is 1 as $s = 0$. An often-used distortion function is $d(x_i, y_j) = (y_j - x_i)^2$. For this distortion function, $\exp(sd_{ij})$ is a Gaussian function (without the coefficient). Therefore, $\exp(sd_{ij})$ can be treated as a truth function or a DCF $T(\theta_{xi}|y)$, where $\theta_{xi}$ is a fuzzy set on $\mathcal{Y}$ instead of $\mathcal{X}$; $\lambda_i$ can be treated as the logical probability $T(\theta_{xi})$ of $x_i$. Now, we can find that Equation (27) is actually a semantic Bayes formula (in Bayes' Theorem III). An $R(D)$ function can be expressed by the semantic mutual information formula with a truth function that is equal to an $R(\Theta)$ function (see Appendix B for details).

In statistical mechanics, a similar distribution to the above $P(y|x_i)$ is the Boltzmann distribution [44]

$$P(x_i|T) = \exp\left(-\frac{e_i}{kT}\right) / Z, \quad Z = \sum_i \exp\left(-\frac{e_i}{kT}\right), \quad (28)$$

where $P(x_i|T)$ is the probability of a particle in the $i$th state with energy $e_i$, or the density of particles in the $i$th state with energy $e_i$; $T$ is the absolute temperature, $k$ is the Boltzmann constant, and $Z$ is the partition function.

Suppose that $e_i$ is the $i$th energy, $G_i$ is the number of states with $e_i$, and $G$ is the total number of all states. Then, $P(x_i) = G_i/G$ is the prior distribution. Hence, the above formula becomes

$$P(x_i|T) = P(x_i) \exp\left(-\frac{e_i}{kT}\right) / Z', \quad Z' = \sum_i P(x_i) \exp\left(-\frac{e_i}{kT}\right). \quad (29)$$

Now, we can find that $\exp[-e_i/(kT)]$ can be treated as a truth function or a DCF, $Z'$ as a logical probability, and Equation (29) as a semantic Bayes formula in Bayes' Theorem III.

For local equilibrium systems, the different areas of a system have different temperatures. I ([14], pp. 102–103) derived the relationship between minimum Shannon mutual information $R(\Theta)$ and thermodynamic entropy $S$ with the help of the truth function or the DCF (see Appendix C for details):

$$R(\Theta) = \ln G - S/(kT). \quad (30)$$

This formula indicates that the maximum entropy principle is equivalent to the minimum mutual information principle.

**4. How the P–T probability Framework and the G Theory Support Popper's Thought**

*4.1. How Popper's Thought about Scientific Progresses is Supported by the Semantic Information Measure*

As early as 1935, in *The Logic of Scientific Discovery* [7], Popper put forward that the less the logical probability of a hypothesis, the easier it is to be falsified, and hence the more empirical information there is. He says:

> "The amount of empirical information conveyed by a theory, or its empirical content, increases with its degree of falsifiability." (p. 96)

> "The logical probability of a statement is complementary to its degree of falsifiability: it increases with decreasing degree of falsifiability. "(p. 102)

However, Popper had not provided a proper semantic information formula. In 1948, Shannon proposed the famous information theory with statistical probability [3]. In 1950, Carnap and Bar-Hillel proposed a semantic information measure with logical probability:

$$I_p = \log(1/m_p), \quad (31)$$

where $p$ is a proposition, $m_p$ is its logical probability. However, this formula is irrelevant to the instance that may or may not make $p$ true. Therefore, it can only indicate how severe the test is, not how well $p$ survives the test.

In 1963, Popper published his book *Conjectures and Refutations* [45]. In this book, he affirms more clearly that the significance of scientific theories is to convey information. He says:

> "It characterizes as preferable the theory which tells us more; that is to say, the theory which contains the greater amount of experimental information or content; which is logically stronger; which has greater explanatory and predictive power; and which can therefore be more severely tested by comparing predicted facts with observations. In short, we prefer an interesting, daring, and highly informative theory to a trivial one." ([45], p. 294).

In this book, Popper proposes using ([45], p. 526)

$$P(e, hb)/P(e, b) \text{ or } \log[P(e, hb)/P(e, b)]$$

to represent how well a theory survives a severe test, where $e$ is interpreted as the supporting evidence of the theory $h$, and $b$ is background knowledge. $P(e, hb)$ and $P(e, b)$ in [45] are conditional probabilities, whose modern forms are $P(e|h, b)$ and $P(e|b)$. Different from Carnap and Bar-Hillel's semantic information formula, the above formula can reflect how well evidence $e$ supports theory $h$. However, $P(e, hb)$ is the conditional probability of $e$. It is not easy to explain $P(e, hb)$ as a logical probability. Hence, we cannot say that $\log[P(e, hb)/P(e, b)]$ represents the amount of semantic information.

There have been several semantic or generalized information measures [38,46,47]. My semantic information measure supports Popper's theory about hypothesis evaluation more than others. We use an example to show its properties.

Suppose that $y_j$ is a GPS pointer or a hypothesis "$x$ is about $x_j$" with a Gaussian truth function $\exp[-(x - x_j)^2/(2\sigma^2)]$ (see Figure 5). Then, we have the amount of semantic information:

$$I(x_i; \theta_j) = \log \frac{P(x_i | \theta_j)}{P(x_i)} = \log \frac{T(\theta_j | x_i)}{T(\theta_j)} = \log[1/T(\theta_j)] - (x_i - x_j)^2/(2\sigma^2). \quad (32)$$

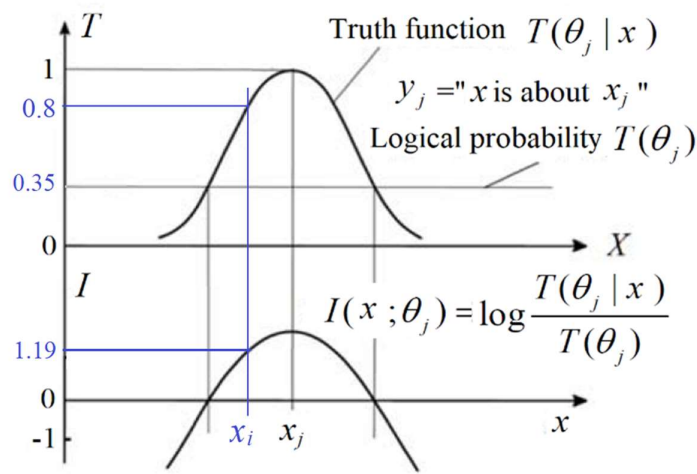

**Figure 5.** Semantic information conveyed by $y_j$ about $x_i$ can represent the verisimilitude of $x_j$ reflecting $x_i$. When real $x$ is $x_i$, the truth value is $T(\theta_j|x_i) = 0.8$; information $I(x_i; \theta_j)$ is $\log(0.8/0.35) = 1.19$ bits. If $x$ exceeds a certain range, the information is negative.

Figure 5 shows that the truth function and the semantic information change with $x$. Small $P(x_i)$ means that $x_i$ is unexpected; large $P(x_i|\theta_j)$ means that the prediction is correct; $\log[P(x_i|\theta_j)/P(x_i)]$ indicates how severe and how well $y_j$ is tested by $x_i$. Large $T(\theta_j|x_i)$ means that $y_j$ is true or close to the truth; small $T(\theta_j)$ means that $y_j$ is precise. Hence, $\log[T(\theta_j|x_i)/T(\theta_j)]$ indicates the verisimilitude of $x_j$ reflecting $x_i$. Unexpectedness, correctness, testability, truth, precision, verisimilitude, and deviation are all reconciled in the formula.

Figure 5 indicates that the less the logical probability is, the more information there is; the larger the deviation is, the less information there is; a wrong hypothesis will convey negative information. According to this measure, the information conveyed by a tautology or a contradiction is 0. These conclusions accord with Popper's thought.

*4.2. How the Semantic Information Measure Supports Popper's Falsification Thought*

Popper affirms that a hypothesis with less logical probability is more easily falsified. If it survives tests, it can convey more information. The semantic information formula for $I(x; \theta_j)$ reflects this thought. Popper affirms that a counterexample can falsify a universal hypothesis. The generalized KL information (Equation (19)) supports this point of view. The truth function of a universal hypothesis only takes value 0 or 1. If there is an instance $x_i$ that makes $T(\theta_j|x_i) = 0$, then $I(x_i; \theta_j)$ is $-\infty$. The average information $I(X; \theta_j)$ is also $-\infty$, which falsifies the universal hypothesis.

Lakatos partly accepted Kuhn's thought about falsification. He pointed out [48] that in scientific practices, scientists do not readily give up a theory (or a hypothesis) when minor observed facts falsify it. They often add some auxiliary hypotheses to the theory or its hard core so that the theory with the auxiliary hypotheses accords with more observed facts. Lakatos hence describes his improved falsification as sophisticated falsification.

Lakatos is correct. For example, according to the hard core of science, distances from a GPS device to three satellites ascertain a GPS device's position. However, to provide accurate positioning, the manufacturers also need to consider additional factors, including satellite geometry, signal blockage, atmospheric conditions, and so on. However, Popper is not necessarily wrong. We can plead for Popper with the following reasons:

- Popper claims that scientific knowledge grows by repeating conjectures and refutations. Repeating conjectures should include adding auxiliary hypotheses.
- Falsification is not the aim. Falsifiability is only the demarcation criterion of scientific and non-scientific theories. The aim of science is to predict empirical facts with more information. Scientists hold a scientific theory depending on if it can convey more information than other theories. Therefore, being falsified does not means being given up.

However, there is still a problem. That is, how can a falsified hypothesis convey more information than others? According to the average semantic information formula or the Generalized KL formula, we can raise the average semantic information by

- increasing the fuzziness or decreasing the predictive precision of the hypothesis to a proper level and
- reducing the degree of belief in a rule or a major premise.

We may change a universal hypothesis into a universal hypothesis that is not strict. For example, we may change "all swans are white" into "almost all swans are white," "all swans may be white," or "swans are white; the degree of confirmation is 0.9".

A GPS device is a quantitative example. A GPS device uses RMS or the like to indicate its accuracy (correctness and precision). A larger circle around the GPS pointer on a GPS map (see Figure 4) means a larger RMS and lower accuracy. If a GPS device shows a few more actual positions beyond the circle, but not too far, we may not give up the GPS device. We can assume that it has a slightly larger RMS and continue to use it. We choose the best hypothesis, as we want a GPS device with the highest accuracy. If a GPS device may direct wrong directions, we may reduce the degree of belief in it to decrease average information loss [12].

*4.3. For Verisimilitude: To Reconcile the Content Approach and the Likeness Approach*

In Popper's earlier works, he emphasizes the importance of hypotheses with less logical probabilities, without stressing hypotheses' truth. In contrast, in Popper's later book *Realism and the Aim of Science* [49], he explains that science aims at better explanatory theories, which accord with facts and hence are true, even if we do not know or cannot prove that they are true. Lakatos [50], therefore, points out that Popper's game of science and his aim of science are contradictive. The game of science consists of bold conjectures and severe refutations; the conjectures need not be true. However, the aim of science consists in developing true or truth-like theories about the mind-independent world. Lakatos' criticism is related to Popper's concept of verisimilitude.

Popper ([42], p. 535) provides a formula for verisimilitude:

$$Vs(a) = \begin{cases} 0, & \text{if } a \text{ is a tautology,} \\ -1, & \text{if } a \text{ is a contradiction,} \\ [1-P(a)]/[1+P(a)], & \text{otherwise,} \end{cases} \quad (33)$$

where $a$ is a proposition, $1 - P(a)$ means the information content. This formula means that the less the logical probability, the more the information content there is, and hence, the higher the verisimilitude is. Popper wishes that $Vs(a)$ changes from −1 to 1. However, according to this formula, $Vs(a)$ cannot appear between −1 and 0. The more serious problem is that he only makes use of logical probability $P(a)$ without using the truth value of proposition $a$ or without using the consequence that may be closer to truth than another [51], so that it cannot express likeness between the prediction and the consequence. Popper later admitted this mistake and emphasized that we need true explanatory theories that accord with the real world [49].

Now, researchers use three approaches to interpret verisimilitude [52]: the content approach, the consequence approach, and the likeness approach. As the latter two are relevant and different from the content approach, theories of verisimilitude have routinely been classified into two rival camps: the content approach and the likeness approach [52]. The content approach emphasizes tests' severity, unlike the likeness approach that emphasizes hypotheses' truth or closeness to truth. Some researchers think that the content approach and the likeness approach are irreconcilable [53], just as Lakatos thinks that Popper's game of science and his aim of science are contradictive. There are also researchers, such as Oddie [52], who try to combine the content approach and the likeness approach. However, they admit that it is not easy to combine them.

This paper continues Oddie's effort. Using the semantic information measure as verisimilitude or truth-likeness, we can combine the content approach and the likeness approach easily.

In Equation (32) and Figure 5, the truth function $T(\theta_j|x)$ is also the confusion probability function; it reflects likeness between $x$ and $x_j$. The $x_i$ (or $X = x_i$) is the consequence, and the distance between $x_i$ and $x_j$ in the feature space reflects the likeness. The $\log[1/T(\theta_j)]$ represents the testing severity and potential information content. Using Equation (32), we can easily explain an often-mentioned example: why "the sun has 9 satellites" (8 is true) has higher verisimilitude than "the sun has 100 satellites" [52].

Another often-mentioned example is how to measure the verisimilitude of weather predictions [52]. Using the semantic information method, we can assess more practical weather predictions. We use a vector $\mathbf{x} = (h, r, w)$ to denote weather, where $h$ is temperature, $r$ is rainfall, and $w$ is wind speed. Let $\mathbf{x}_j = (h_j, r_j, w_j)$ be the predicted weather and $\mathbf{x}_i = (h_i, r_i, w_i)$ be the actual weather (consequence). The prediction is $y_j$ = "$\mathbf{x}$ is about $\mathbf{x}_j$". For simplicity, we assume that $h$, $r$, and $w$ are independent. The Gaussian truth function may be:

$$T(\theta_j|\mathbf{x}) = \exp[-(h-h_j)^2/(2\sigma_h^2) - (r-r_j)^2/(2\sigma_r^2) - (w-w_j)^2/(2\sigma_w^2)]. \quad (34)$$

This truth function accords with the core of the likeness approach [52]: that the likeness of a proposition depends on the distance between two instances (**x** and **x**$_j$). If the consequence is **x**$_i$, then the truth value $T(\theta_j|\mathbf{x}_i)$ of proposition $y_j(\mathbf{x}_i)$ is the likeness. Additionally, the information $I(\mathbf{x}_i; \theta_j) = \log[T(\theta_j|\mathbf{x}_i)/T(\theta_j)]$ is the verisimilitude, which has almost all desirable properties for which the three approaches are used.

This verisimilitude $I(\mathbf{x}_i; \theta_j)$ is also related to prior probability distribution $P(\mathbf{x})$. The correct prediction of unusual weather has much higher verisimilitude than that of common weather if both predictions are right.

Now, we can explain that the Regularized Least Squared Error criterion is getting popular, because it is similar to the maximum average verisimilitude criterion (refer to Equation (21)).

**5. The P–T Probability Framework and the G Theory Are Used for Confirmation**

*5.1. The Purpose of Confirmation: Optimizing the Degrees of Belief in Major Premises for Uncertain Syllogisms*

Confirmation is often explained as assessing the evidential impact on hypotheses (incremental confirmation) [11,54] or the evidential support for hypotheses (absolute confirmation) [55,56]. The degree of confirmation is the degree of belief that is supported by evidence or data [57]. Researchers only use confirmation measures to assess hypotheses. There have been various confirmation measures [11]. I also proposed two different confirmation measures [58]. In the following, I briefly introduce my study on confirmation related to scientific reasoning.

In my study on confirmation, the task, purpose, and method of confirmation are a little different from those in the popular confirmation theories.

- The task of confirmation:

  Only major premises, such as "if the medical test is positive, then the tested person is infected" and "if $x$ is a raven, then $x$ is black", need confirmation. The degrees of confirmation are between −1 and 1. A proposition, such as "Tom is elderly", or a predicate, such as "$x$ is elderly" ($x$ is one of the given people), needs no confirmation. The truth function of the predicate reflects the semantic meaning of "elderly" and is determined by the definition or the idiomatic usage of "elderly". The degree of belief in a proposition is a truth value, and that in a predicate is a logical probability. The truth value and the logical probability are between 0 and 1 instead of −1 and 1.

- The purpose of confirmation:

  The purpose of confirmation is not only for assessing hypotheses (major premises), but also for probability predictions or uncertain syllogisms. A syllogism needs a major premise. However, as pointed out by Hume and Popper, it is impossible to obtain an absolutely right major premise for an infinite universe by induction. However, it is possible to optimize the degree of belief in the major premise by the proportions of positive examples and counterexamples. The optimized degree of belief is the degree of confirmation. Using a degree of confirmation, we can make an uncertain or fuzzy syllogism. Therefore, confirmation is an important link in scientific reasoning according to experience.

- The method of confirmation:

  I do not directly define a confirmation measure, as most researchers do. I derive the confirmation measures by optimizing the degree of belief in a major premise with the maximum semantic information criterion or the maximum likelihood criterion. This method is also the method of statistical learning, where the evidence is a sample.

From the perspective of statistical learning, a major premise comes from a classifier or a predictive rule. We use the medical test as an example to explain the relationship between classification and confirmation (see Figure 6). The binary signal detection and the classification with labels "elder" and "non-elder" are similar. Classifications for watermelons and junk emails are also similar.

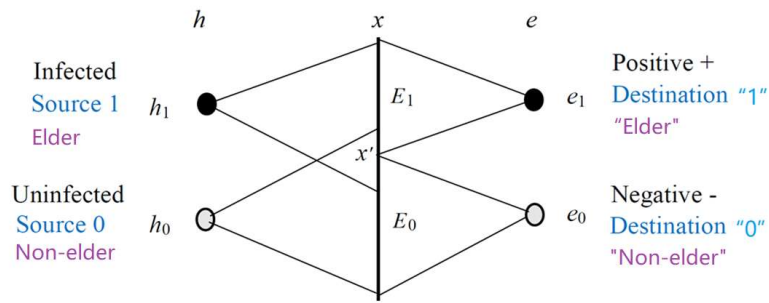

**Figure 6.** Illustrating the medical test and the binary classification for explaining confirmation. If $x$ is in $E_1$, we use $e_1$ as prediction "$h$ is $h_1$"; if $x$ is in $E_0$, we use $e_0$ as prediction "$h = h_0$".

In Figure 6, $h_1$ denotes an infected specimen (or person), $h_0$ denotes an uninfected specimen, $e_1$ is positive, and $e_0$ is negative. We can treat $e_1$ as a prediction "$h$ is infected" and $e_0$ as a prediction "$h$ is uninfected". The $x$ is the observed feature of $h$; $E_1$ and $E_2$ are two sub-sets of the universe of $x$. If $x$ is in $E_1$, we select $e_1$; if $x$ is in $E_0$, we select $e_0$. For the binary signal detection, we use "0" or "1" in the destination to predict 0 or 1 in the source according to the received analog signal $x$.

The two major premises to be confirmed are "if $e_1$ then $h_1$", denoted by $e_1 \rightarrow h_1$, and "if $e_0$ then $h_0$", denoted by $e_0 \rightarrow h_0$. A confirmation measure is denoted by $c(e \rightarrow h)$.

The $x'$ ascertains a classification. For a given classification or predictive rule, we can obtain a sample including four examples ($e_1$, $h_1$), ($e_0$, $h_1$), ($e_1$, $h_0$), and ($e_0$, $h_0$). Then, we can use the four examples' numbers $a$, $b$, $c$, and $d$ (see Table 1) to construct confirmation measures.

**Table 1.** The numbers of four types of examples for confirmation measures.

|  | $e_0$ | $e_1$ |
|---|---|---|
| $h_1$ | b | a |
| $h_0$ | d | c |

The $a$ is the number of example ($e_1$, $h_1$). The $b$, $c$, and $d$ are in like manner. An absolute confirmation measure can be expressed as function $f(a, b, c, d)$. Its increment is

$$\Delta f = f(a + \Delta a, b + \Delta b, c + \Delta c, d + \Delta d) - f(a, b, c, d).$$

Information measures and confirmation measures are used for different tasks. To compare two hypotheses and choose a better one, we use an information measure. Using the maximum semantic information criterion, we can find the best $x'$ for the maximum mutual information classifications [12]. To optimize the degree of belief in a major premise, we need a confirmation measure. An information measure is used for classifications, whereas a confirmation measure is used after classifications.

*5.2. Channel Confirmation Measure b\* for Assessing a Classification as a Channel*

A binary classification ascertains a Shannon channel, which includes four conditional probabilities, as shown in Table 2.

**Table 2.** The four conditional probabilities of a binary classification form a Shannon channel.

|  | $e_0$ (Negative) | $e_1$ (Positive) |
|---|---|---|
| $h_1$ (Infected) | $P(e_0|h_1) = b/(a + b)$ | $P(e_1|h_1) = a/(a + b)$ |
| $h_0$ (Uninfected) | $P(e_0|h_0) = d/(c + d)$ | $P(e_1|h_0) = c/(c + d)$ |

We regard predicate $e_1(h)$ as the combination of believable and unbelievable parts (see Figure 7). The truth function of the believable part of $e_1$ is $T(E_1|h) \in \{0,1\}$. There are $T(E_1|h_1) = T(E_0|h_0) = 1$ and $T(E_1|h_0) = T(E_1|h_0) = 0$. The unbelievable part is a tautology, whose truth function is always 1. Then, we have the truth functions of predicates $e_1(h)$ and $e_0(h)$:

$$T(\theta_{e1}|h) = b_1' + b_1 T(E_1|h); \tag{35}$$

$$T(\theta_{e0}|h) = b_0' + b_0 T(E_0|h). \tag{36}$$

Model parameter $b_1$ is the proportion of the believable part, and $b_1' = 1 - |b_1|$ is the proportion of the unbelievable part and also the truth value of $y_1(h_0)$, where $h_0$ is a counter-instance. The $b_1'$ may be regarded as the degree of disbelief in the major premise $e_1 \rightarrow h_1$.

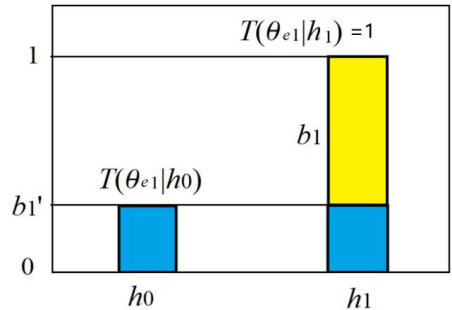

**Figure 7.** Truth function $T(\theta_{e1}|h)$ includes the believable part with proportion $b_1$ and the unbelievable part with proportion $b_1'$ ($b_1' = 1 - |b_1|$).

The four truth values form a semantic channel, as shown in Table 3.

**Table 3.** The semantic channel ascertained by two degrees of disbelief $b_1'$ and $b_0'$.

|  | $e_0$ **(Negative)** | $e_1$ **(Positive)** |
|---|---|---|
| $h_1$ (infected) | $T(\theta_{e0}|h_1) = b_0'$ | $T(\theta_{e1}|h_1) = 1$ |
| $h_0$ (uninfected) | $T(\theta_{e0}|h_0) = 1$ | $T(\theta_{e1}|h_0) = b_1'$ |

Using $T(\theta_{e1}|h)$, we can make probability prediction $P(h|\theta_{e1}) = P(h)T(\theta_{e1}|h)/T(\theta_{e1})$. According to the generalized KL formula (Equation (19)), when $P(h|\theta_{e1}) = P(h|e_1)$ or $T^*(\theta_{e1}|h) \propto P(e_1|h)$, the average semantic information $I(h; \theta_{e1})$ reaches its maximum. Letting $P(h_1|\theta_{e1}) = P(h_1|e_1)$, we derive (see Appendix D for details):

$$b_1^* = 1 - b_1'^* = [P(e_1|h_1) - P(e_1|h_0)]/P(e_1|h_1). \tag{37}$$

Considering $P(h_1|e_1) < P(h_0|e_1)$, we have

$$b_1^* = b_1'^* - 1 = [P(e_1|h_0) - P(e_1|h_1)]/P(e_1|h_0). \tag{38}$$

Combining the above two formulas, we have

$$b_1^* = b^*(e_1 \rightarrow h_1) = \frac{P(e_1|h_1) - P(e_1|h_0)}{\max(P(e_1|h_1), P(e_1|h_0))} = \frac{ad - bc}{\max(a(c+d), c(a+b))} = \frac{LR^+ - 1}{\max(LR^+, 1)}, \tag{39}$$

where *LR* is likelihood ratio, and $LR^+$ is positive *LR*. Elles and Fitelson [54] proposed Hypothesis Symmetry: $c(e_1 \rightarrow h_1) = -c(e_1 \rightarrow h_0)$. As we also have $c(h_1 \rightarrow e_1) = -c(h_1 \rightarrow e_0)$, where two consequents are opposite, I called this symmetry Consequence Symmetry [58]. As

$$b_1^* = b^*(e_1 \rightarrow h_0) = \frac{P(e_1|h_0) - P(e_1|h_1)}{\max(P(e_1|h_0), P(e_1|h_1))} = -b^*(e_1 \rightarrow h_1), \tag{40}$$

$b_1^*$ possesses this symmetry.

In like manner, we obtain

$$b_0^* = b^*(e_0 \to h_0) = \frac{P(e_0|h_0) - P(e_0|h_1)}{\max(P(e_0|h_0), P(e_0|h_1))} = \frac{LR^- - 1}{\max(LR^-, 1)} \quad (41)$$

Using the above symmetry, we have $b^*(e_1 \to h_0) = -b^*(e_1 \to h_1)$ and $b^*(e_0 \to h_1) = -b^*(e_0 \to h_0)$.

In the medical test, the likelihood ratio measure *LR* is used for assessing how reliable a testing result (positive or negative) is. Measure *F* proposed by Kemeny and Oppenheim [55] is

$$F(e_1 \to h_1) = \frac{P(e_1|h_1) - P(e_1|h_0)}{P(e_1|h_1) + P(e_1|h_0))} = \frac{LR^+ - 1}{LR^+ + 1}. \quad (42)$$

Like measure *F*, measure $b^*$ is also the function of *LR*, and hence both can be used for assessing the medical test. Compared with *LR*, $b^*$ and *F* can indicate the distance between a test (any $b^*$) and the best test ($b^* = 1$) or the worst test ($b^* = -1$) better.

Compared with *F*, $b^*$ is better for probability predictions. For example, from $b_1^* > 0$ and $P(h)$, we have

$$P(h_1|\theta_{e1}) = P(h_1)/[P(h_1) + b_1^* P(h_0)] = P(h_1)/[1 - b_1^* P(h_0)]. \quad (43)$$

This formula is simpler than the classical Bayes formula (see Equation (6)). If $b_1^* = 0$, then $P(h_1|\theta_{e1}) = P(h_1)$. If $P(h_1|\theta_{e1}) < 0$, then we can make use of Consequent Symmetry to make the probability prediction [58]. So far, it is still problematic to use $b^*$, *F*, or another measure to assess how well a probability prediction or clarify the Raven Paradox.

*5.3. Prediction Confirmation Measure $c^*$ for Clarifying the Raven Paradox*

Statistics not only uses the likelihood ratio to indicate how reliable a testing method (as a channel) is, but also uses the correct rate to indicate how possible the predicted event is. Measures *F* and $b^*$, like *LR*, cannot indicate the quality of a probability prediction. For example, $b_1^* > 0$ does not mean $P(h_1|\theta_{e1}) > P(h_0|\theta_{e1})$. Most other confirmation measures have similar problems [58].

We now treat probability prediction $P(h|\theta_{e1})$ as the combination of a believable part with proportion $c_1$ and an unbelievable part with proportion $c_1'$, as shown in Figure 8. We call $c_1$ the degree of belief in rule $e_1 \to h_1$ as a prediction.

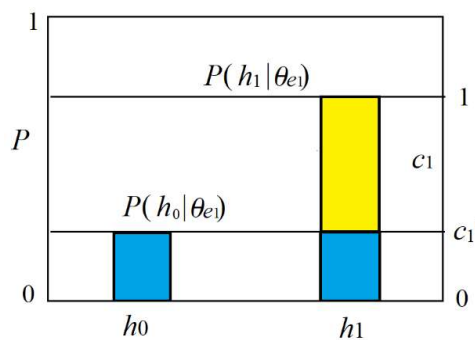

**Figure 8.** Likelihood function $P(h|\theta_{e1})$ may be regarded as a believable part plus an unbelievable part.

When the prediction accords with the fact, e.g., $P(h|\theta_{e1}) = P(h|e_1)$, $c_1$ becomes $c_1^*$. Then, we derive the prediction confirmation measure

$$c_1^* = c^*(e_1 \to h_1) = \frac{P(h_1|e_1) - P(h_0|e_1)}{\max(P(h_1|e_1), P(h_0|e_1))}$$
$$= \frac{2P(h_1|e_1) - 1}{\max(P(h_1|e_1), 1 - P(h_1|e_1))} = \frac{2CR_1 - 1}{\max(CR_1, 1 - CR_1)}. \quad (44)$$

where $CR_1 = P(h_1|\theta_{e1}) = P(h_1|e_1)$ is the correct rate of rule $e_1 \to h_1$. Letting both the numerator and denominator of Equation (44) multiply by $P(e_1)$, we obtain

$$c_1^* = c^*(e_1 \to h_1) = \frac{P(h_1, e_1) - P(h_0, e_1)}{\max(P(h_1, e_1), P(h_0, e_1))} = \frac{a-c}{\max(a,c)}. \quad (45)$$

In like manner, we obtain

$$c_0^* = c^*(e_0 \to h_0) = \frac{P(h_0, e_0) - P(h_1, e_0)}{\max(P(h_0, e_0), P(h_1, e_0))} = \frac{d-b}{\max(d,b)}. \quad (46)$$

It is easy to prove that $c^*(e_1 \to h_1)$ also possesses Consequence Symmetry. Making use of this symmetry, we can obtain $c^*(e_1 \to h_0) = -c^*(e_1 \to h_1)$ and $c^*(e_0 \to h_1) = -c^*(e_0 \to h_0)$.

When $c_1^* > 0$, according to Equation (45), we have the correct rate of rule $e_1 \to h_1$:

$$CR_1 = P(h_1|\theta_{e1}) = 1/(1+c_1'^*) = 1/(2-c_1^*). \quad (47)$$

When $c^*(e_1 \to h_1) < 0$, we may make use of Consequence Symmetry to make the probability prediction. However, when $P(h)$ is changed, we should still use $b^*$ with $P(h)$ for probability predictions.

For the medical test, we need both conformation measures. Measure $b^*$ tells the reliability of a test as means or the channel in comparison with other tests, whereas measure $c^*$ tells the possibility that a person is infected. For scientific predictions, such as earthquake predictions, $b^*$ and $c^*$ have similar meanings.

Now, we can use measure $c^*$ to clarify the Raven Paradox.

Hemple [59] proposed the confirmation paradox or the Raven Paradox. According to the Equivalence Condition in the classical logic, "if $x$ is a raven, then $x$ is black" (Rule I) is equivalent to "if $x$ is not black, then $x$ is not a raven" (Rule II). A piece of white chalk supports Rule II; hence it also supports Rule I. However, according to the Nicod Criterion [60], a black raven supports Rule I, a non-black raven undermines Rule I, and a non-raven thing, such as a black cat or a piece of white chalk, is irrelevant to Rule I. Hence, there exists a paradox between the Equivalence Condition and the Nicod Criterion.

To clarify the Raven Paradox, some researchers, including Hemple [59], affirm the Equivalence Condition and deny the Nicod Criterion; some researchers, such as Scheffler and Goodman [61], affirm the Nicod Criterion and deny the Equivalence Condition. Some researchers do not fully affirm or deny the Equivalence Condition or the Nicod Criterion.

Figure 9 shows a sample for the major premise "if $x$ is a raven, then $x$ is black".

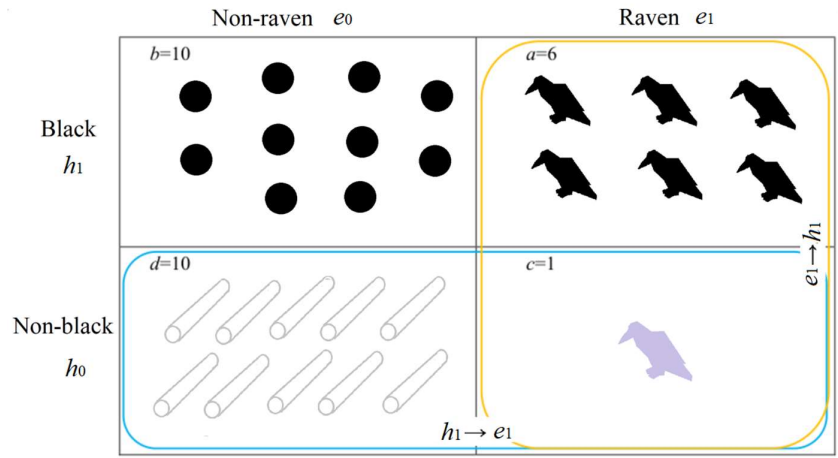

Figure 9. Using a sample to examine different confirmation measures for the Raven Paradox. Source: author.

First, we consider measure $F$ to see if we can use it to eliminate the Raven Paradox. The difference between $F(e_1 \rightarrow h_1)$ and $F(h_0 \rightarrow e_0)$ is that their counterexamples are the same ($c = 1$), yet their positive examples are different. When $d$ increases to $d + \Delta d$, $F(e_1 \rightarrow h_1) = (ad - bc)/(ad + bc + 2ac)$ and $F(h_0 \rightarrow e_0) = (ad - bc)/(ad + bc + 2dc)$ unequally increase. Therefore, though measure $F$ denies the Equivalence Condition, it still affirms that $\Delta d$ affects both $F(e_1 \rightarrow h_1)$ and $F(h_0 \rightarrow e_0)$, and hence, measure $F$ does not accord with the Nicod Criterion.

Among all confirmation measures, only measure $c^*$ can ensure that $f(a, b, c, d)$ is only affected by $a$ and $c$. There is $c^*(e_1 \rightarrow h_1) = (6 - 1)/6 = 5/6$.

However, many researchers still think that the Nicod Criterion is incorrect; this criterion accords with our intuition only because a confirmation measure $c(e_1 \rightarrow h_1)$ can evidently increase with $a$ and slightly increase with $d$. For example, Fitelson and Hawthorne [62] believe that measure $LR$ may be used to explain that a black raven can confirm "ravens are black" more strongly than a non-black non-raven thing. Is it true?

For the above example, $LR$, $F$, and $b^*$ do ensure that the increment $\Delta f$ caused by $\Delta a = 1$ is bigger than $\Delta f$ caused by $\Delta d = 1$. However, when $a = d = 20$ and $b = c = 10$, except for $c^*$, no measure can ensure $\Delta f/\Delta a > \Delta f/\Delta d$ (see Table 13 in [58] for details), which means that all popular confirmation measures cannot be used to explain that a black raven can confirm "ravens are black" more strongly than a piece of white chalk.

As $c^*(e_1 \rightarrow h_1) = (a - c)/\max(a, c)$ and $c^*(h_0 \rightarrow e_0) = (d - c)/\max(d, c)$, the Equivalence Condition does not hold, and measure $c^*$ accords with the Nicod Criterion very well. Therefore, the Raven Paradox does not exist anymore according to measure $c^*$.

*5.4. How Confirmation Measures F, b\*, and c\* are Compatible with Popper's Falsification Thought*

Popper affirms that a counterexample can falsify a universal hypothesis or a major premise. However, for an uncertain major premise, how do counterexamples affect its degree of confirmation? Confirmation measures $F$, $b^*$, and $c^*$ can reflect that the existence of fewer counterexamples is more important than the existence of more positive examples.

Popper affirms that a counterexample can falsify a universal hypothesis, according to which we can explain that for the falsification of a strict universal hypothesis, it is important to have no counterexample. Now, for the confirmation of a major premise that is not strict universal hypothesis, we can explain that it is important to have fewer counterexamples. Therefore, confirmation measures $F$, $b^*$, and $c^*$ are compatible with Popper's falsification thought.

Scheffler and Goodman [61] proposed selective confirmation based on Popper's falsification thought. They believe that black ravens (whose number is $a$) support "ravens are black", because black ravens undermine "ravens are not black". Their reason why non-black ravens (whose number is $c$) support "ravens are not black" is that non-black ravens undermine the opposite hypothesis "ravens are black". Their explanation is significant. However, they did not provide the corresponding confirmation measure. Measure $c^*(e_1 \rightarrow h_1)$ is what they need.

Now, we can find that it is confirmation that allows us not to give up a falsified major premise and to keep it with an optimized degree of belief for obtaining more information.

More discussions about confirmation can be found in [58].

## 6. Induction, Reasoning, Fuzzy Syllogisms, and Fuzzy Logic

*6.1. Viewing Induction from the New Perspective*

From the perspective of statistical learning, induction includes:

- induction for probability predictions: to optimize likelihood functions with sampling distributions,
- induction for the semantic meanings or the extensions of labels: to optimize truth functions with sampling distributions, and

- induction for the degrees of confirmation of major premises: to optimize the degrees of belief in major premises with the proportions of positive examples and counterexamples after classifications.

From the above perspective, induction includes confirmation, and confirmation is part of induction. Only rules or if-then statements with two predicates need confirmation.

The above inductive methods also include the selections or conjectures of predictive models (likelihood functions and truth functions) with the semantic information criterion. The iteration algorithms for mixture models and maximum mutual information classifications [12] reflect repeating conjectures and refutations. Therefore, these inductive methods are compatible with Popper's thought that scientific progress comes from conjectures and refutations.

*6.2. The Different Forms of Bayesian Reasoning as Syllogisms*

According to the above analyses, Bayesian reasoning has different forms, as shown in Table 4. The interpretations with blue words indicate newly added contents because of the use of the P–T probability framework.

Table 4. The different forms of Bayesian reasoning.

| Reasoning between/with | Major Premise or Model | Minor Premise or Evidence | Consequence | Interpretation |
|---|---|---|---|---|
| Between two instances | $P(y_j\|x)$ | $x_i$ ($X = x_i$) | $P(y_j\|x_i)$ | Conditional SP |
| | | $y_j$, $P(x)$ | $P(x\|y_j) = P(x)P(y_j\|x)/P(y_j)$ $P(y_j) = \sum_i P(y_j\|x_i) P(x_i)$ | Bayes' Theorem II (Bayes' prediction) |
| Between two sets | $T(\theta_2\|\theta)$ | $y_1$ (is true) | $T(\theta_2\|\theta_1)$ | Conditional LP |
| | | $y_2$, $T(\theta)$ | $T(\theta\|\theta_2) = T(\theta_2\|\theta)T(\theta)/T(\theta_2)$, $T(\theta_2) = T(\theta_2\|\theta)T(\theta) + T(\theta_2\|\theta')T(\theta')$ | Bayes' Theorem I ($\theta'$ is the complement of $\theta$) |
| Between an instance and a set (or model) | $T(\theta_j\|x)$ | $X = x_i$ or $P(x)$ | $T(\theta_j\|x_i)$ or $T(\theta_j) = \sum_i T(\theta_j\|x_i)P(x_i)$ | Truth value and logical probability |
| | | $y_j$ is true, $P(x)$ | $P(x\|\theta_j) = P(x)T(\theta_j\|x)/T(\theta_j)$, $T(\theta_j) = \sum_i T(\theta_j\|x_i)P(x_i)$ | The semantic Bayes prediction in Bayes' Theorem III |
| | $P(x\|\theta_j)$ | $x_i$ or $\mathbf{D}_j$ | $P(x_i\|\theta_j)$ or $P(\mathbf{D}_j\|\theta_j)$ | Likelihood |
| | | $P(x)$ | $T(\theta_j\|x) = [P(x\|\theta_j)/P(x)]/\max[P(x\|\theta_j)/P(x)]$ | Inference in Bayes' Theorem III |
| Induction: with sampling distributions to train predictive models | $P(x\|\theta_j)$ | $P(x\|y_j)$ | $P^*(x\|\theta_j)$ (optimized $P(x\|\theta_j)$ with $P(x\|y_j)$) | Likelihood Inference |
| | $P(x\|\theta)$ and $P(\theta)$ | $P(x\|y)$ or $\mathbf{D}$ | $P(\theta\|\mathbf{D}) = P(\theta)P(\mathbf{D}\|\theta)/\sum_j P(\theta_j)P(\mathbf{D}\|\theta_j)$ | Bayesian Inference |
| | $T(\theta_j\|x_j)$ | $P(x\|y_j)$ and $P(x)$ | $T^*(\theta_j\|x) = P(x\|y_j)/P(x)/\max[P(x\|y_j)/P(x)]$ $= P(y_j\|x)/\max[P(y_j\|x)]$ | Logical Bayesian Inference |
| With degree of channel confirmation | $b_1^* = b^*(e_1 \to h_1) > 0$ | $h$ or $P(h)$ | $T(\theta_{e1}\|h_1) = 1$, $T(\theta_{e1}\|h_0) = 1 - \|b_1^*\|$; $T(\theta_{e1}) = P(h_1) + (1 - b_1^*)P(h_0)$ | Truth values and logical probability |
| | | $e_1$ ($e_1$ is true), $P(h)$ | $P(h_1\|\theta_{e1}) = P(h_1)/T(\theta_{e1})$, $P(h_0\|\theta_{e1}) = (1 - b_1^*)P(h_0)/T(\theta_{e1})$; | A fuzzy syllogism with $b^*$ |
| With degree of prediction confirmation | $c_1^* = c^*(e_1 \to h_1) > 0$ | $e_1$ ($e_1$ is true) | $P(h_1\|\theta_{e1}) = 1/(2 - c_1^*)$, $P(h_0\|\theta_{e1}) = (1 - c_1^*)/(2 - c_1^*)$ | A fuzzy syllogism with $c^*$ |

*SP is statistical probability; LP is logical probability. Blue words indicate the new forms of Bayesian reasoning.

It is important that the reasoning with truth functions or degrees of confirmation is compatible with statistical reasoning (using Bayes' Theorem II). When $T^*(\theta_j|x) \propto P(y_j|x)$ or $T(\theta_{ej}|h) \propto P(e_j|h)$, the consequences are the same as those from the classical statistics.

On the other hand, the fuzzy syllogisms are compatible with the classical syllogism, because when $b_1^* = 1$ or $c_1^* = 1$, for given $e_1$, the consequence is $P(h_1) = 1$ and $P(h_0) = 0$.

However, the above fuzzy syllogisms have their limitations. The syllogism with $b^*(e_1 \rightarrow h_1)$ is the generalization of a classical syllogism. The fuzzy syllogism is:

- The major premise is $b^*(e_1 \rightarrow h_1) = b_1^*$
- The minor premise is $e_1$ with $P(h)$,
- The consequence is $P(h|\theta_{e1}) = P(h_1)/[P(h_1) + (1 - b_1^*)P(h_0)] = P(h_1)/[1 - b_1^*P(h_0)]$.

When $b_1^* = 1$, if the minor premise becomes "$x$ is in $E_2$, and $E_2$ is included in $E_1$", then this syllogism becomes Barbara (AAA-1) [63], and the consequence is $P(h_1|\theta_{e1}) = 1$. Hence, we can use the above fuzzy syllogism as the fuzzy Barbara.

As the Equivalence Condition does not hold for both channels' confirmation and predictions' confirmation, if the minor premise is $h = h_0$, we can only use a converse confirmation measure $b^*(h_0 \rightarrow e_0)$ or $c^*(h_0 \rightarrow e_0)$ as the major premise to obtain the consequence (see [58] for details).

*6.3. Fuzzy Logic: Expectations and Problems*

Truth functions are also membership functions. Fuzzy logic can simplify statistics for membership functions and truth functions.

Suppose that the membership functions of three fuzzy sets $A$, $B$, and $C$ are $a(x)$, $b(x)$, and $c(x)$, respectively. Let the logical expression of $A$, $B$, and $C$ be $F(A, B, C)$ with three operators $\cap$, $\cup$, and $^c$ and the logical expression of three truth functions be $f(a(x), b(x), c(x))$ with three operators $\wedge$, $\vee$, and $^-$. The operators $\cap$ and $\wedge$ can be omitted. There are $2^8$ different expressions with $A$, $B$, and $C$. To simplify statistics, we expect

$$T(F(A, B, C)|x) = f(a(x), b(x), c(x)). \tag{48}$$

If the above equation is tenable, we only need to optimize three truth functions $a(x)$, $b(x)$, and $c(x)$ and calculate the truth functions of various expressions $F(A, B, C)$. Further, we expect that the fuzzy logic in $f(...)$ is compatible with Boolean algebra and Kolmogorov's probability system.

However, in general, the logical operations of membership functions do not follow Boolean algebra. Zadeh's fuzzy logic is defined with [19]:

$$a(x) \wedge b(x) = \min(a(x), b(x)),$$
$$a(x) \vee b(x) = \max(a(x), b(x)), \tag{49}$$
$$\overline{a(x)} = 1 - a(x).$$

According to this definition, the law of complementarity does not hold (e.g., $A \cap A^c \neq \phi$ and $A \cup A^c \neq \mathcal{X}$) since

$$a(x) \wedge \overline{a(x)} = \min(a(x), 1 - a(x)) \neq 0,\ a(x) \vee \overline{a(x)} = \max(a(x), 1 - a(x)) \neq 1.$$

There are also other definitions. To consider correlation between two predicates "$x$ is in $A$" and "$x$ is in $B$", Wang et al. [64] define:

$$a(x) \wedge b(x) = \begin{cases} \min(a(x), b(x)), & \text{positively correlated,} \\ a(x)b(x), & \text{independent,} \\ \max(0, a(x)+b(x)-1), & \text{negatively correlated.} \end{cases} \tag{50}$$

$$a(x) \vee b(x) = \begin{cases} \max(a(x), b(x)), & \text{positively correlated,} \\ a(x) + b(x) - a(x)b(x), & \text{independent,} \\ \min(1, a(x)+b(x)), & \text{negatively correlated.} \end{cases} \tag{51}$$

If two predicates are always positively correlated, then the above operations become Zadeh's operations. According the above definition, since $A$ and $A^c$ are negatively correlated, there are

$$a(x) \wedge \overline{a(x)} = \max(0, a(x) + 1 - a(x) - 1) \equiv 0,\ a(x) \vee \overline{a(x)} = \min(1, a(x) + 1 - a(x)) \equiv 1.$$

Thus, the law of complementarity is tenable.

To build a symmetrical model of color vision in the 1980s, I [65,66] proposed the fuzzy quasi-Boolean algebra on $\mathcal{X}$, which is defined with the Boolean algebra on cross-set [0,1]×$\mathcal{X}$. This model of color vision is explained as a fuzzy 3–8 decoder, and hence is called the decoding model (see Appendix E). Its practicability can be proved by the fact that the International Commission on Illumination (CIE) recommended a symmetrical model, which is almost the same as mine, for color transforms in 2006 [67]. Using the decoding model, we can easily explain color blindness and color evolution by spitting or merging three sensitivity curves [68].

We can use the fuzzy quasi-Boolean algebra to simplify statistics for the truth functions of natural languages. In these cases, we need to distinguish atomic labels and compound labels and assume that the random sets for the extensions of these atomic labels become wider or narrower at the same time.

Suppose that the truth functions of three atomic labels "youth", "adult", and "elder" are $u(x)$, $a(x)$, and $e(x)$, respectively. The truth functions of three compound labels are

$$T(\text{"child"}|x) = T(\text{"non-youth" and "non-adult"}|x) = [\overline{u}\,\overline{a}] = 1 - \max(u(x), a(x)),$$

$$T(\text{"youth" and "non-adult"}|x) = [u\,\overline{a}] = \max(0, u(x) - a(x)),$$

$$T(\text{"midlle age"}|x) = T(\text{"adult" and "non-youth" and "non-elder"}|x)$$

$$= [a\,\overline{u}\,\overline{e}] = \max(0, a(x) - \max(u(x), e(x))).$$

However, the correlation between different labels or statements is often complicated. When we choose a set of operators of fuzzy logic, we need to balance reasonability and feasibility.

**7. Discussions**

*7.1. How the P–T Probability Framework has Been Tested by Its Applications to Theories*

The P–T Probability Framework includes statistical probabilities and logical probabilities. Logical probabilities include truth functions, which are also membership functions and can reflect the extensions of hypotheses or labels. Using truth function $T(\theta_j|x)$ and prior distribution $P(x)$, we can produce likelihood function $P(x|\theta_j)$ and train $T(\theta_j|x)$ and $P(x|\theta_j)$ with sampling distributions $P(x|y_j)$. Then, we can let a machine reason like the human brain with the extensions of concepts.

By semantic information methods with this framework, we can use logical probabilities to express semantic information: $I(x_i; \theta_j) = \log[T(\theta_j|x)/T(\theta_j)]$. We can also use the likelihood function to express predictive information, because $I(x_i; \theta_j) = \log[T(\theta_j|x)/T(\theta_j)] = \log[P(x_i|\theta_j)/P(x_i)]$. When we calculate the average semantic information $I(X; \theta_j)$ and $I(X; \Theta)$, we also need statistical probabilities, such as $P(x|y_j)$, to express sampling distributions.

In the popular methods of statistical learning, only statistical probabilities, including subjective probabilities, are used. For binary classification, we can use a pair of logistic functions. However, for multi-label learning and classification, there is no simple method [40]. With (fuzzy) truth functions now, we can easily solve the extensions of multi-labels for classifications by Logical Bayesian Inference [12]. Using Shannon's channels with statistical probabilities and semantic channels with logical probabilities, we can let two channels mutually match to achieve maximum mutual information classifications and to speed up the convergence of mixture models with better convergence proofs [12].

Section 3.3 shows that the semantic Bayes formula with the truth function and the logical probability has been used in rate-distortion theory and statistical mechanics. Therefore, the P–T probability framework can be used not only for communication or epistemology but also for control or ontology.

For the evaluation of scientific theories and hypotheses, with logical probabilities, we can use semantic information $I(x; \theta_j)$ as verisimilitude and testing severity, both of which can be mutually converted. With statistical probabilities, we can express how sampling distributions test hypotheses.

For confirmation, using truth functions, we can conveniently express the degrees of belief in major premises. With statistical probabilities, we can derive the degrees of confirmation, e.g., the degrees of belief optimized by sampling distributions.

For Bayesian reasoning, with the P–T probability framework, we can have more forms of reasoning (see Table 4), such as:

- two types of reasoning with Bayes' Theorem III,
- Logical Bayesian Inference from sampling distributions to optimized truth functions, and
- fuzzy syllogisms with the degrees of confirmation of major premises.

In short, with the P–T probability framework, we can resolve many problems more conveniently.

*7.2. How to Extend Logic to the Probability Theory?*

Jaynes [16] concludes that the probability theory is the extension of logic. This conclusion is a little different from that probability is the extension of logic. The former includes the latter. The former emphasizes that probability is an important tool for scientific reasoning, which includes Bayesian reasoning, the maximum likelihood estimation, the maximum entropy method, and so on. Although Jaynes interprets probabilities with average truth values ([16], p. 52), this interpretation is not the extension. When an urn contains $N$ balls with only two colors: white and red, we may use truth value $R_i$ to represent a red ball on the $i$th draw and $\overline{R}_i = W_i$ to represent a white ball on the $i$th draw. However, if the balls have more colors, this interpretation will not sound so good. In this case, the frequentist interpretation is simpler.

Zadeh's fuzzy set theory is an essential extension of the classical logic to probability for logical probability. Using (fuzzy) truth functions or membership functions, we can let a machine learn human brains' reasoning with fuzzy extensions. This reasoning can be an essential supplement to the probabilistic reasoning that Jaynes summarized.

With Kolmogorov's axiomatic system, some researchers wish to develop probability logic with logical expression $f(T(A), T(B), …)$ so that

$$T(F(A, B, …)) = f(T(A), T(B),…). \quad (52)$$

This task is difficult, because it is hard to avoid the contradiction between the probability logic and statistical reasoning. According to Kolmogorov's axiomatic system, we have

$$T(A \cup B) = T(A) + T(B) − T(A \cap B). \quad (53)$$

However, we cannot obtain $T(A \cap B)$ from $T(A)$ and $T(B)$, because $T(A \cap B)$ is related to $P(x)$ and the correlation between two predicates "$x$ is in $A$" and "$x$ is in $B$". The operators "∧" and "∨" can only be used for the operations of truth values instead of logical probabilities. For example, $A$ is a set {youths}, and $B$ is a set {adults}. Suppose that when people include high school students and their parents, the prior distribution is $P_1(x)$; when people include high school students and soldiers, the prior distribution is $P_2(x)$. $T_1(A)$ is obtained from $P_1(x)$ and $T(A|x)$, $T_2(A)$ from $P_2(x)$ and $T(A|x)$, and so on. $T_1(A \cap B)$ should be much smaller than $T_2(A \cap B)$, even if $T_1(A) = T_2(A)$ and $T_1(B) = T_2(B)$.

A better method for $T(A \cap B)$ than Equation (52) is first to obtain truth function, such as $T(A \cap B|x) = T(A|x) \wedge T(B|x)$. Then we have

$$T(A \cap B) = \sum_i P(x_i) T(A \cap B | x_i). \quad (54)$$

Similarly, to obtain $T(F(A, B, …))$, we should first obtain truth function $f(T(A|x), T(B|x), …)$ using fuzzy logic operations.

Various forms of Bayesian reasoning in Table 4 for the extension of the classical logic are compatible with statistical reasoning. For example, Bayes' Theorem III is compatible with Bayes' Theorem II. If we use a probability logic for a fuzzy or uncertain syllogism, we had better check if the reasoning is compatible with statistical reasoning.

In the two types of probability logic proposed by Reichenbach [9] and Adams [26], there is

$$P(p => q) = 1 - P(p) + P(pq), \tag{55}$$

as the extension of $p => q$ ($p$ implies $q$) in mathematical logic, where $p$ and $q$ are two proposition sequences or predicates. My extension is different. I used confirmation measures $b^*(e_1 \to h_1)$ and $c^*(e_1 \to h_1)$, which may be negative, to represent the extended major premises. The most extended syllogisms in Table 4 are related to Bayes' formulas. The measure $c^*(p \to q)$ is the function of $P(q|p)$ (see Equation (44)); they are compatible. It should also be reasonable to use $P(q|p)$ ($p$ and $q$ are two predicates) as the measure for assessing a fuzzy major premise. However, $P(q|p)$ and $P(p => q)$ are different. We can prove

$$P(q|p) \leq P(p => q) = 1 - P(p) + P(pq). \tag{56}$$

The proof can be found in Appendix F. $P(p => q)$ may be much bigger than $P(q|p)$. For example, if $P(p) = 0.1$ and $P(pq) = 0.02$, then $P(q|p) = 0.2 < P(p => q) = 0.92$.

Equation (55) is used because there is $(p \Rightarrow q) = \overline{p} \vee q = \overline{p} \vee pq$ in mathematical logic. However, no matter how many examples support $p => q$, the degree of belief in "if $p$ then $q$" cannot be 1, as pointed by Hume and Popper [7]. When counterexamples exist, $p => q$ becomes $P(p => q) = 1 - P(p) + P(pq)$, which is also not suitable as the measure for assessing fuzzy major premises.

In addition, it is much simpler to obtain $P(q)$ or $P(pq)$ from $P(q|p)$ and $P(p) = 1$ than from $P(p => q)$ and $P(p) = 1$ [9,26]. If $P(p) < 1$, according to statistical reasoning, we also need $P(q|\overline{p})$ to obtain $P(q) = P(p)P(q|p) + [1 - P(p)]P(q|\overline{p})$. Using a probability logic, we should be careful about whether the results are compatible with statistical reasoning.

*7.3. Comparing the Truth Function with Fisher's Inverse Probability Function*

In the P–T probability framework, the truth function plays an important role.

It is called Likelihood Inference to use the likelihood function $P(x|\theta_j)$ as the inference tool (where $\theta_j$ is a constant). It is called Bayesian Inference to use Bayesian posterior $P(\theta|\mathbf{D})$ as the inference tool (where $P(\theta|\mathbf{D})$ means parameters' posterior distribution for given data or a sample). It is called the Logical Bayesian Inference to use the truth function $T(\theta_j|x)$ as the inference tool [12].

Fisher called the parameterized TPF $P(\theta_j|x)$ as inverse probability [5]. As $x$ is a variable, we had better call $P(\theta_j|x)$ the Inverse Probability Function (IPF). According to Bayes' Theorem II, there are

$$P(\theta_j|x) = P(\theta_j) P(x|\theta_j)/P(x), \tag{57}$$

$$P(x|\theta_j) = P(x_i) P(\theta_j|x)/P(\theta_j). \tag{58}$$

IPF $P(\theta_j|x)$ can make use of the prior knowledge $P(x)$ well. When $P(x)$ is changed, $P(\theta_j|x)$ can still be used for probability predictions. However, why did Fisher and other researchers give up $P(\theta_j|x)$ as the inference tool?

When $n = 2$, we can easily construct $P(\theta_j|x)$, $j = 1,2$, with parameters. For instance, we can use a pair of logistic functions as the IPFs. Unfortunately, when $n > 2$, it is hard to construct $P(\theta_j|x)$, $j = 1,2, \ldots, n$, because there is normalization limitation $\sum_j P(\theta_j|x) = 1$ for every $x$. That is why a multi-class or multi-label classification is often converted into several binary classifications [40].

$P(\theta_j|x)$ and $P(y_j|x)$ as predictive models also have a serious disadvantage. In many cases, we can only know $P(x)$ and $P(x|y_j)$ without knowing $P(\theta_j)$ or $P(y_j)$ so that we cannot obtain $P(y_j|x)$ or $P(\theta_j|x)$. Nevertheless, we can get truth function $T^*(\theta_j|x)$ in these cases. There is no normalization limitation, and hence it is easy to construct a group of truth functions and train them with $P(x)$ and $P(x|y_j)$, $j = 1, 2, \ldots, n$, without $P(y_j)$ or $P(\theta_j)$.

We summarize that the truth function has the following advantages:

- We can use an optimized truth function $T^*(\theta_j|x)$ to make probability prediction for different $P(x)$ as well as we use $P(y_j|x)$ or $P(\theta_j|x)$.
- We can train a truth function with parameters by a sample with a small size as well as we train a likelihood function.

- With the truth function, we can indicate the semantic meaning of a hypothesis or the extension of a label.
  It is also the membership function, which is suitable for classification.
- To train a truth function $T(\theta_j|x)$, we only need $P(x)$ and $P(x|y_j)$, without needing $P(y_j)$ or $P(\theta_j)$.
- Letting $T^*(\theta_j|x) \propto P(y_j|x)$, we can bridge statistics and logic.

*7.4. Answers to Some Questions*

If probability theory is the logic of science, may any scientific statement be expressed by a probabilistic or uncertain statement? The P–T probability framework supports Jaynes' point of view. The answer is yes. The Gaussian truth function can be used to express a physical quantity with a small uncertainty, as well as it is used to express the semantic meaning of a GPS pointer. The semantic information measure can be used to evaluate the result predicted by a physical formula. Guo [69] studied the operations of fuzzy numbers and concluded that the expectation of the function of fuzzy numbers is equal to the function of the expectations of the fuzzy numbers. For example, (fuzzy 3)*(fuzzy 5) = (fuzzy 15) = (fuzzy 3*5). According to this conclusion, the probabilistic or uncertain statements may be compatible with certain statements about physical laws.

Regarding the question of whether the existing probability theory, the P–T probability framework, and the G theory are scientific theories and if they can be tested by empirical fact. My answer is that they are not scientific theories tested by empirical facts but tools for scientific theories and applications. They can be tested by their abilities to resolve problems with scientific theories and applications.

Regarding the question of whether there are identifiable logical constraints for assessing scientific theories: I believe that a scientific theory needs not only self-consistency, but also compatibility with other accepted scientific theories. Similarly, a unified theory of probability and logic as a scientific tool should also be compatible with other accepted mathematical theories, such as statistics. For this reason, in the P–T probability framework, an optimized truth function (or a semantic channel) is equivalent to a TPF (or a Shannon channel) when we use them for probability predictions. Additionally, the semantic mutual information has its upper limit: Shannon's mutual information; the fuzzy logic had better be compatible with Boolean algebra.

*7.5. Some Issues that Need Further Studies*

Many researchers are talking about interpretable AI, but the problem of meaning is still very much with us today [70]. The P–T probability framework should be helpful for interpretable AI. The reasons are:

- The human brain thinks using the extensions (or denotations) of concepts more than interdependencies. A truth function indicates the (fuzzy) extension of a label and reflects the semantic meaning of the label; Bayes' Theorem III expresses the reasoning with the extension.
- The new confirmation methods and the fuzzy syllogisms can express the induction and the reasoning with degrees of belief that the human brain uses, and the reasoning is compatible with statistical reasoning.
- The Boltzmann distribution has been applied to the Boltzmann machine [71] for machine learning. With the help of the semantic Bayes formula and the semantic information methods, we can better understand this distribution and the Regularized Least Square criterion related to information.

However, it is still not easy to interpret neural networks with semantic information for machine learning. Can we interpret a neural network as a semantic channel that consists of a set of truth functions? Can we apply the Channels' Matching algorithm [12] to neural networks for maximum mutual information classifications? These issues need further studies.

This paper provides a bridge between statistics and logic. It should be helpful for the semantic dictionary based on statistics that AI needs. However, there is much work to do, such as to design and optimize the truth functions of terms in natural languages, such as "elder", "heavy rain",

"normal body temperature", and so on. The difficulty is that the extensions of these terms change from area to area. For example, the extension of "elder" depends on the life span of people in the area; the extensions on rainfalls of "heavy rain" in coastal areas and in desert areas are different. These extensions are related to prior distribution $P(x)$. For unifying logical methods and statistical methods for AI, the efforts of more people are needed.

When samples are not big enough, the degrees of confirmation we obtain are not reliable. In these cases, we need to combine the hypothesis-testing theory to replace the degrees of confirmation with the degree intervals of confirmation. We need further study.

This paper only extends a primary syllogism with the major premise "if $e = e_1$ then $h = h_1$" to some effective fuzzy syllogisms. It will be complicated to extend more syllogisms [63]. We need further study for the extension.

We may use truth functions as DCFs and the generalized information/entropy measures for random events' control and statistical mechanics (see Section 3.3). We need further studies for practical results.

## 8. Conclusions

As pointed by Popper, a hypothesis has both statistical probability and logical probability. This paper has proposed the P–T probability framework that uses "$P$" to denote statistical probability and "$T$" to denote logical probability. In this framework, the truth function of a hypothesis is equivalent to the membership function of a fuzzy set. Using the new Bayes theorem (Bayes' Theorem III), we can convert a likelihood function and a truth function one to another so that we can use sampling distributions to train truth functions. The maximum semantic information criterion used is equivalent to the maximum likelihood criterion. Statistics and logic are hence connected.

I have introduced how the P–T probability framework is used for the semantic information G theory or the G theory. The G theory is a natural generalization of Shannon's information theory. It can be used to improve statistical learning and explain the relationship between information and thermodynamic entropy, in which the minimum mutual information distribution is equivalent to the maximum entropy distribution.

I have shown how the P–T probability framework and the G theory support Popper's thought about scientific progress, hypothesis evaluation, and falsification. The semantic information measure can reflect Popper's testing severity and verisimilitude. The semantic information approach about verisimilitude can reconcile the content approach and the likeness approach [52].

I have shown how to use the semantic information measure to derive channel confirmation measure $b^*$ and prediction confirmation measure $c^*$. Measure $b^*$ is compatible with the likelihood ratio and changes between -1 and 1. It can be used to assess medical tests, signal detections, and classifications. Measure $c^*$ can be used to assess probability predictions and clarify the Raven Paradox. Two confirmation measures are compatible with Popper's falsification thought.

I have provided several different forms of Bayesian reasoning, including fuzzy syllogisms with confirmation measures $b^*$ and $c^*$. I have introduced a fuzzy logic that was used to set up a symmetrical model of color vision. This fuzzy logic is compatible with Boolean algebra, and hence compatible with the classical logic.

The above theoretical applications of the P–T probability framework illustrate its reasonability and practicability. We should be able to find the wider applications of the P–T probability framework. However, to combine the logical and statistical methods for AI, there is still much work to do. In order to apply the P–T probability framework and the G theory to deep learning with neural networks and to random events' control for practical results, we need further studies.

**Funding:** This research received no external funding.

**Acknowledgments:** I thank the editors of *Philosophies* and its special issue *Science and Logic* for giving me a chance to introduce this study systematically. I also thank the two reviewers for their comments, which vastly improve this paper. I should also thank people who influenced my destiny so that I have technical and theoretical research experience for this article.

**Conflicts of Interest:** The author declares no conflict of interest.

### Appendix A. The Proof of Bayes' Theorem III

**As** joint probability $P(x, \theta_j) = P(X = x, X \in \theta_j) = P(x|\theta_j)T(\theta_j) = T(\theta_j|x)P(x)$, there are

$$P(x|\theta_j) = P(x)T(\theta_j|x)/T(\theta_j), \quad T(\theta_j|x) = T(\theta_j)P(x|\theta_j)/P(x). \tag{A1}$$

As $P(x|\theta_j)$ is horizontally normalized, $T(\theta_j)$ in Equation (8) is $\sum_i P(x_i) T(\theta_j|x_i)$. As $T(\theta_j|x)$ is longitudinally normalized, we have

$$1 = \max[T(\theta_j)P(x|\theta_j)/P(x)] = T(\theta_j)\max[P(x|\theta_j)/P(x)]. \tag{A2}$$

Hence $T(\theta_j) = 1/\max[P(x|\theta_j)/P(x)]$.

### Appendix B. A $R(D)$ Function is Equal to a $R(\Theta)$ Function with Truth Functions

When $T(\theta_{xi}|y) = \exp(sd_{ij})$, we have

$$\begin{aligned} R(\Theta) &= \min_{P(x),\Theta_x} I(X;Y) = I(Y;\Theta_x) = \sum_i P(x_i) \sum_j P(y_j|x_i) \log \frac{T(\theta_{xi}|y_j)}{T(\theta_{xi})} \\ &= \sum_i P(x_i) \sum_j P(y_j|x_i) \log \frac{\exp(sd_{ij})}{\lambda_i} \\ &= sD(s) - \sum_i P(x_i) \log \lambda_i = \min_{P(x),D} I(X;Y) = R(D). \end{aligned} \tag{A3}$$

### Appendix C. The Relationship between Information and Thermodynamic Entropy [14]

Helmholtz free energy is

$$F = E - TS. \tag{A4}$$

For an equilibrium system,

$$S = E/T - F = E/T - kN \ln Z, \quad Z = \sum_i G_i \exp[-e_i/(kT)] \tag{A5}$$

where $N$ is the number of particles. For a local equilibrium system,

$$S = \sum_j \frac{E_j}{T_j} + k \sum_j N_j \ln Z_j, \tag{A6}$$

where $T_j$ is the temperature of the $j$th area ($y_j$), and $N_j$ is the number of particles in the $j$th area. We now consider minimum mutual information $R(\Theta)$ for given distribution constraint functions $T(\theta_j|x) = \exp[-e_i/(kT_j)]$ ($j = 1, 2, \ldots$). The logical probability of $y_j$ is $T(\theta_j) = Z_j/G$, and the statistical probability is $P(y_j) = N_j/N$. From Appendix B and the above equation for $S$, we derive

$$\begin{aligned} R(\Theta) &= -\sum_j P(y_j) \frac{e_j}{kT_j} - \sum_j P(y_j) \ln(Z_j/G) \\ &= -\sum_j \frac{N_j}{N} \frac{(E_j/N_j)}{kT_j} - \sum_j \frac{N_j}{N} \ln(Z_j/G) = \ln G - S/(kN), \end{aligned} \tag{A7}$$

where $e_j = E_j/N_j$ is the average energy of a particle in the $j$-th area.

### Appendix D. The Derivation for $b_1^*$

Let $P(h_1|\theta_{e1}) = P(h_1|e_1)$. From

$$P(h_1 \mid e_{\theta 1}) = \frac{P(h_1)T(\theta_{e1} \mid h_1)}{T(\theta_{e1})} = \frac{P(h_1)}{P(h_1) + b_1' P(h_0)}, \quad (A8)$$

$$P(h_1 \mid e_1) = \frac{P(h_1)P(e_1 \mid h_1)}{P(h_1)P(e_1 \mid h_1) + P(h_0)P(e_1 \mid h_1)}, \quad (A9)$$

we have $b_1'^* = P(e_1|h_0)/P(e_1|h_1)$ for $P(h_1|e_1) \geq P(h_0|e_1)$. Hence,

$$b_1^* = 1 - b_1'^* = [P(e_1|h_1) - P(e_1|h_0)]/P(e_1|h_1). \quad (A10)$$

**Appendix E. Illustrating the Fuzzy Logic in the Decoding Model of Color Vision**

From cones' responses *b*, *g*, and *r*, we can use the fuzzy decoding to produce the eight outputs as mental color signals, as shown in Figure A11.

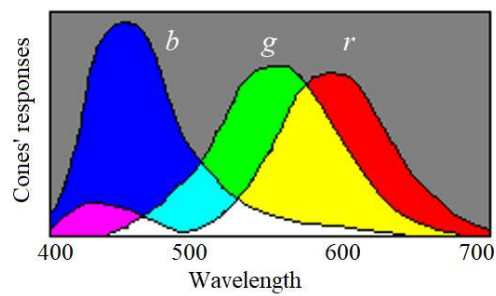

**Figure A1.** Illustrating the fuzzy logic in the decoding model.

**Appendix F. To Prove $P(q|p) \leq P(p \Rightarrow q)$**

Because of $P(p) = P(pq) + P(p\bar{q}) \leq 1$, we have $P(pq) \leq 1 - P(p\bar{q})$, and thus

$$P(q \mid p) = \frac{P(pq)}{P(p)} = \frac{P(pq)}{P(pq) + P(p\bar{q})} \leq \frac{1 - P(p\bar{q})}{1 - P(p\bar{q}) + P(p\bar{q})} \quad (A11)$$
$$= 1 - P(p\bar{q}) = P(\bar{P} \vee Pq) = 1 - P(p) + P(pq) = P(p \Rightarrow q).$$